\documentclass[oupdraft]{biostatistics}
\usepackage[colorlinks=true, urlcolor=citecolor, linkcolor=citecolor, citecolor=citecolor]{hyperref}
\usepackage{threeparttable}
\usepackage{url}

\history{Received xxx, 2013;
revised xxx, 2013;
accepted for publication xxx, 2013}

\begin{document}

\title{Joint Analysis of Differential Gene Expression in Multiple Studies using Correlation Motifs}

\author{YINGYING WEI, HONGKAI JI$^\ast$,\\
[4pt]
\textit{Department of Biostatistics, Johns Hopkins University Bloomberg School of Public Health, Baltimore, Maryland, USA}
\\[2pt]
{hji@jhsph.edu}}

\markboth%
{YY Wei and HK Ji}
{Correlation Motif Discovery}

\maketitle

\footnotetext{To whom correspondence should be addressed.}

\begin{abstract}
{The standard methods for detecting differential gene expression
are mostly designed for analyzing a single gene expression experiment.
When data from multiple related gene expression studies are available, separately
analyzing each study is not an ideal strategy as it may fail to detect
important genes with consistent but relatively weak differential signals
in multiple studies. Jointly modeling all data allows one to borrow information
across studies to improve the analysis. However, a simple concordance
model, in which each gene is assumed to be differential in either all studies or
none of the studies, is incapable of handling genes with study-specific
differential expression. In contrast, a model that naively enumerates and analyzes all possible
differential patterns across all studies can deal with study-specificity and
allow information pooling, but the complexity of its parameter space grows exponentially
as the number of studies increases. Here we propose a {\em correlation motif \/} approach to address this dilemma.
This approach automatically searches for a small number of latent probability
vectors called {\em correlation motifs \/} to capture the major
correlation patterns among multiple studies. The motifs provide
the basis for sharing information among studies and genes. The approach improves
detection of differential expression and overcomes the barrier of exponentially
growing parameter space. It is capable of handling all possible study-specific
differential patterns in a large number of studies. The advantages of this new
approach over existing methods are illustrated using both simulated and real data.}
{Bayes hierarchical model;
Correlation motif; EM algorithm; Microarray; Multiple Datasets.}
\end{abstract}

\section{Introduction}
\label{sec: Introduction}
Detecting differentially expressed genes is a basic task
in the analysis of gene expression data.
The state-of-the-art solutions to this problem, such as {\em limma\/}
(\citealp{limmaref}), {\em SAM\/} (\citealp{rsam}), edgeR (\citealp{redgeR1,redgeR2}) and DESeq (\citealp{rDESeq}),
are mostly designed for analyzing data from a single experiment or study.
With 1,000,000+ samples stored in public databases such as
Gene Expression Omnibus (GEO), it is now very common for scientists to
have data from multiple related experiments or studies. An emerging problem
is how one can integrate data from multiple studies to more effectively analyze
differential expression.

One example that motivated this article is a study of the vertebrate Sonic Hedgehog (SHH)
signaling pathway. SHH is a signaling protein that can bind to PTCH1, a
receptor protein in cell membrane (Figure 1(a)).
PTCH1 can interact with another membrane protein SMO to repress its activity.
In the absence of SHH, PTCH1 keeps SMO inactive.
The presence of SHH will repress PTCH1 and activate SMO.
The active SMO triggers a signaling cascade by modulating activities
of three transcription factors, GLI1, GLI2 and GLI3, which in turn
will induce or repress the expression of hundreds of downstream target genes.
SHH pathway is one of the core signaling pathways in vertebrate development.
It is associated with multiple types of tumors and birth defects (\citealp{rSHH1,rSHH2}).
To elucidate the underlying mechanism linking this pathway to diseases,
multiple studies have been performed in different contexts to identify genes whose transcriptional activities
are modulated by SHH signaling. Some studies perturb the SHH signal in different
tissues by knocking out or over-expressing the pathway's key signal transduction components
such as SHH, PTCH1 and SMO, while others compare disease samples with corresponding controls.
Table 1 contains eight such datasets in mouse originally generated and compiled by \citealp{rSHHAndrew0}
and \citealp{rSHHMao}. Each dataset involves a comparison of genome-wide expression profiles between two different sample types.
These data were all collected using Affymetrix Mouse Expression Set 430 arrays. The questions
of biological interest include (1) which genes are controlled by the SHH signal in each dataset,
(2) which genes are the core targets that respond to the SHH signal irrespective of tissue type and
developmental stage, and (3) which genes are context-specific targets and are modulated by the SHH
signal only in certain conditions. For simplicity, below we will call each dataset a {\em study\/}.

One simple approach to analyze these data is to analyze each study separately
using existing state-of-the-art methods such as {\em limma\/}
(\citealp{limmaref}) or {\em SAM\/} (\citealp{rsam}).
This approach is not ideal as it may fail to detect genes with low fold changes
but consistently differential in many or all studies.

Modeling all data jointly may allow one to borrow information across studies
to improve the analysis. A simple model to combine data is to assume that each gene is either differential in all studies
or non-differential in all studies (\citealp{rConlon}). This concordance model may help with identifying
genes with small but consistent expression changes in all studies. However, it ignores the
reality that activities of many important genes are tissue- or time-specific.
This method will only produce a single gene list that reports and ranks genes in the same way for all studies.
It cannot  prioritize genes differently for different studies to account for context-specificity.

A more flexible approach is to consider all possible differential expression patterns.
Suppose there are $D$ studies and each gene can either be differential or non-differential in each study,
there will be $2^D$ possible differential expression patterns. One can model the data as a mixture of $2^D$
different gene classes. This allows one to deal with context-specificity. However, an obvious drawback
is that as the number of studies increases, the number of possible patterns increases exponentially.
Thus the model does not scale well with the increasing $D$.

In this article, we propose a new method, {\em CorMotif\/}, for jointly analyzing multiple studies to improve
differential expression detection. This method is both flexible for handling context-specificity
and scalable to increasing study number. The key idea is to use a small number of
latent probability vectors called ``correlation motifs'' to model the major correlation patterns among the studies.
The motifs essentially group genes into clusters based on their differential expression patterns,
and the differential gene detection is coupled with the clustering.

Previously, \cite{reb1} proposed a method for analyzing differential expression involving multiple biological conditions.
This method, abbreviated as ``eb1'' hereinafter, requires users to specify all possible differential patterns, and the data are then modeled accordingly.
If a user applies this method to detect differential expression between two conditions in multiple
studies and wants to accommodate all possible differential patterns, the user has to enumerate all $2^D$ possible patterns,
leading to the exponential complexity problem. Similar to \cite{reb1}, \cite{rJen} developed a hierarchical Bayesian
model and a Markov Chain Monte Carlo (MCMC) algorithm to analyze multiple conditions, again with exponential complexity due to
requirement of enumerating all possible patterns.
\cite{reb2011} generalizes \cite{reb1} to a model that can integrate information from multiple studies where each study may involve comparisons of multiple conditions. Within each study, this method enumerates all possible combinatorial patterns among multiple conditions (again exponential complexity). Across
studies, differential expression patterns are assumed to be concordant, that is, each gene is assumed to have the same differential pattern in all studies.
The concordance assumption does not allow study-specific differential expression.

\cite{rXDE} proposed a fully Bayesian framework, XDE, for cross-study differential expression analysis. It offers two implementations. The ``Single-Indicator'' implementation uses a concordance model by assuming that each gene's differential state is the same across all studies. The ``Multiple-Indicator'' implementation allows study-specific differential expression. However, it assumes that all genes have the same prior probability to be differential within the same study, and the differential states of each gene in different studies are a priori independent. Conceptually, these assumptions are similar to a {\em CorMotif\/} model with a single cluster, which often is insufficient to capture the heterogeneity among genes since the cross-study correlation pattern may vary from one gene to another (see details later). XDE does not have the exponential complexity problem, but it uses MCMC for posterior inference and is very slow computationally.

To capture the heterogeneity among genes, \cite{reb10} developed a method for simultaneous clustering and differential expression analysis.
Similar to {\em CorMotif\/}, this method also assumes that genes belong to multiple clusters, and different clusters have different propensities to show differential expression. However, \cite{reb10} only considered detecting differential expression between two conditions in one study. Although one may conceptually
extend this approach to handle multiple studies by combining it with the model developed by \cite{reb1}, such a simple extension
would lead to a model (called ``eb10best'' hereinafter) in which genes are assumed to fall into multiple clusters and each cluster is a mixture of $2^D$ differential patterns. As a result, the complexity of the parameter space would become $O(K*2^D)$ where $K$ is the number of clusters.

In summary, none of the tools discussed above allows one to integrate information from multiple studies and also addresses study-specificity, heterogeneity among genes, and exponential complexity at the same time. These are the issues {\em CorMotif\/} attempts to solve. We organize this article as follows. Section 2 introduces the {\em CorMotif\/} model and algorithm. Section 3 uses simulations to demonstrate the approach. In Section 4, {\em CorMotif\/} will be applied to the SHH data. Section 5 will provide remarks and discussions. Here, we focus on discussing {\em CorMotif\/} for microarray data since it was motivated by the microarray analysis in the SHH study. However, the idea behind {\em CorMotif\/} is general, and it should be straight-forward to develop a similar framework for RNA-seq data.

\section{Methods}
\label{sec2}

\subsection{Data Structure and Preprocessing}
\label{datastructure}
Suppose there are $G$ genes and $D$ microarray studies. Each study $d$ compares two biological conditions (e.g., cancer vs. normal), and each condition $l$ has $n_{dl}$ replicate samples. Different studies may be related, but they can compare different biological conditions. Let $x_{gdlj}$ denote the normalized and appropriately transformed expression value of gene $g$ in study $d$, condition $l$ and replicate $j$. In this article, all microarray data were normalized and log-transformed using RMA (\citealp{rRMA}). The collection of all observed data is

\noindent $\mbox{\boldmath{$X$}}=\left\{x_{gdlj}: g=1,\ldots,G;  d=1,\ldots,D;  l=1,2;  j=1,\ldots,n_{dl}\right\}.$

Each gene can be differentially expressed in some, all, or none of the studies. Let $a_{gd}=1$ or $0$ indicate whether gene $g$ is differentially expressed in study $d$ or not. $\mbox{\boldmath{$A$}} = (a_{gd})_{G \times D}$ is a $G \times D$ matrix that contains all $a_{gd}$s. Given the observed data $\mbox{\boldmath{$X$}}$, one is interested in inferring $\mbox{\boldmath{$A$}}$.

{\em CorMotif\/} first applies limma \citep{limmaref} to each study separately. Define $\bar{x}_{gdl}=\sum_{j} x_{gdlj} / n_{dl}$, $n_d = n_{d1}+n_{d2}$ and $v_{d}=\frac{1}{n_{d1}}+\frac{1}{n_{d2}}$. For gene $g$ and study $d$, compute the mean expression difference $y_{gd}=\bar{x}_{gd1}-\bar{x}_{gd2}$ and sample variance $s_{gd}^2 = \sum_l \sum_j (x_{gdlj}-\bar{x}_{gdl})^2 / (n_d-2)$. The limma approach assumes that $y_{gd}$s and $s_{gd}^2$s within each study $d$ follow a hierarchical model: (1) $[y_{gd} | \mu_{gd}, \sigma_{gd}^2] \sim N(\mu_{gd}, v_{d} \sigma_{gd}^2)$, (2) $\mu_{gd}=0$ if $a_{gd}=0$, (3) $[\mu_{gd} | a_{gd}=1, \sigma_{gd}^2] \sim N(0, w_{d} \sigma_{gd}^2)$, (4) [$s_{gd}^2 | \sigma_{gd}^2] \sim \frac{\sigma_{gd}^2}{n_d-2}\chi^2_{n_d-2}$, and (5) $[\frac{1}{\sigma_{gd}^2}] \sim \frac{1}{n_{0d}s_{0d}^2} \chi^2_{n_{0d}}$. Here $w_d$, $n_{0d}$ and $s_{0d}^2$ are unknown parameters. Their values can be estimated using the procedure described in \cite{limmaref}. This hierarchical model allows one to pool information across genes to stabilize the variance estimates. \cite{limmaref} shows that it can significantly improve differential gene detection when the sample size $n_d$ is small.
For each study $d$, limma produces a moderated t-statistic for each gene $g$, computed as $t_{gd}=y_{gd}/\sqrt{v_{d}\tilde{s}_{gd}^2}$ where $\tilde{s}_{gd}^2 = \frac{n_{0d}s_{0d}^2+(n_d-2)s_{gd}^2}{n_{0d}+n_d-2}$. This statistic summarizes gene $g$'s differential expression information in study $d$. Under this model, when gene $g$ is not differentially expressed in study $d$ (i.e., $a_{gd}=0$), $t_{gd}$ follows a t-distribution $t_{n_{0d}+n_d-2}$; when $a_{gd}=1$, $t_{gd}$ follows a scaled t-distribution $(1+w_{d}/v_d)^{1/2}t_{n_{0d}+n_d-2}$ \citep{limmaref}.

Next, we arrange all $t_{gd}$s into a matrix $\mbox{\boldmath{$T$}} = (t_{gd})_{G \times D}$. {\em CorMotif\/} will then use $\mbox{\boldmath{$T$}}$ instead of the raw expression values $\mbox{\boldmath{$X$}}$ to infer $\mbox{\boldmath{$A$}}$.

\subsection{Correlation Motif Model}

Organize the differential expression states of gene $g$ into a vector $\mbox{\boldmath{$a$}}_g=[a_{g1},a_{g2},\cdots,a_{gD}]$. For $D$ studies, $\mbox{\boldmath{$a$}}_g$ has $2^D$ possible configurations. A simple way to describe the correlation among studies is to document the empirical frequency of observing each of the $2^D$ configurations of $\mbox{\boldmath{$a$}}_g$ among all genes. This is because $f(\mbox{\boldmath{$a$}}_g)$,
the joint distribution of $[a_{g1},a_{g2},\cdots,a_{gD}]$, is known once the probability of observing each configuration is given. This joint distribution will determine how $a_{gd}$s from different studies are correlated. While simple, this approach is not scalable since it requires $O(2^D)$ parameters and the parameter space expands exponentially with increasing $D$.

To avoid this limitation, {\em CorMotif\/} adopts a hierarchical mixture model (Figure 1(b)). The model assumes that genes fall into $K$ different classes ($K \ll 2^D$), and the moderated t-statistics $\mbox{\boldmath{$T$}} = (t_{gd})_{G \times D}$ are viewed as generated as follows.

\begin{itemize}
\item First, each gene $g$ is randomly and independently assigned a class label $b_g$ according to probability
$\mbox{\boldmath{$\pi$}}=(\pi_1,...,\pi_K)$. Here, $\pi_k \equiv Pr(b_g=k)$ is the prior probability that a gene belongs to class $k$, and $\sum_k \pi_k = 1$.

\item Second, given genes' class labels (i.e., $b_g$s), genes' differential expression states $a_{gd}$s are generated independently according to probabilities
$q_{kd} \equiv Pr(a_{gd}=1|b_g=k)$. For genes in the same class $k$, $\mbox{\boldmath{$a$}}_g$s are generated using the same probabilities $\mbox{\boldmath{$q_{k}$}} = (q_{k1},...,q_{kD})$.

\item Third, given the differential expression states $a_{gd}$s, genes' moderated t-statistics $t_{gd}$s are generated independently according to $f_{d1}(t_{gd})=f(t_{gd}|a_{gd}=1) \sim (1+w_{d}/v_d)^{1/2}t_{n_{0d}+n_d-2}$ or $f_{d0}(t_{gd})=f(t_{gd}|a_{gd}=0) \sim t_{n_{0d}+n_d-2}$.
\end{itemize}

Let $\mbox{\boldmath{$B$}} = (b_{1},...,b_{G})$ be the class membership for all genes. Organize $\mbox{\boldmath{$q_{k}$}}$ into a matrix $\mbox{\boldmath{$Q$}}=(\mbox{\boldmath{$q_{1}^T$}},\cdots, \mbox{\boldmath{$q_{K}^T$}})^T= (q_{kd})_{K \times D}$. Let $\delta(\cdot)$ be an indicator function: $\delta(\cdot)=1$ if its argument is true, and $\delta(\cdot)=0$ otherwise.
Based on the above model, the joint probability distribution of $\mbox{\boldmath{$A$}}$, $\mbox{\boldmath{$B$}}$ and $\mbox{\boldmath{$T$}}$ conditional on $\mbox{\boldmath{$\pi$}}$ and $\mbox{\boldmath{$Q$}}$ is:
\begin{equation}
Pr(\mbox{\boldmath{$T$}},\mbox{\boldmath{$A$}},\mbox{\boldmath{$B$}}|\mbox{\boldmath{$\pi$}},\mbox{\boldmath{$Q$}})=\prod_{g=1}^{G}\prod_{k=1}^{K}\{\pi_k\prod_{d=1}^D[q_{kd}f_{d1}(t_{gd})]^{a_{gd}}[(1-q_{kd})f_{d0}(t_{gd})]^{1-a_{gd}}\}^{\delta{(b_g=k)}}
\end{equation}

According to this model, each gene class $k$ is associated with a vector $\mbox{\boldmath{$q_{k}$}}$ whose elements are the prior probabilities of a gene in this class to be differential in studies $1,\ldots,D$. Each $\mbox{\boldmath{$q_{k}$}}$ represents a probabilistic differential expression pattern and therefore is called a ``motif''. Since $q_{kd}$s are probabilities, genes in the same class can have different $\mbox{\boldmath{$a$}}_g$ configurations. On the other hand, genes from the same class share the same $\mbox{\boldmath{$q_{k}$}}$, and hence their differential expression configuration $\mbox{\boldmath{$a$}}_g$s tend to be similar. Genes in different classes have different $\mbox{\boldmath{$q_{k}$}}$s, and their $\mbox{\boldmath{$a$}}_g$s also tend to be different. Essentially, our model groups genes into $K$ clusters based on $\mbox{\boldmath{$a$}}_g$. However, unlike an usual clustering algorithm, here $\mbox{\boldmath{$a$}}_g$s are unknown.

Despite the assumption that $a_{gd}$s are a priori independent conditional on the class label $b_g$, $a_{gd}$s are no longer independent once the class label $b_g$ is integrated out. To see this, consider the prior probability that a gene is differentially expressed in all studies. Based on our model, $Pr(\mathbf{a}_{g}=[1,\cdots,1])=\sum_{k}(\pi_k\prod_d{q_{kd}})$. A priori, the probability for a gene to be differential in study $d$ is $Pr(a_{gd}=1)=\sum_{k}\pi_kq_{kd}$. If $a_{gd}$s from different studies are independent, one would expect $Pr(\mathbf{a}_{g}=[1,\cdots,1])=\prod_d Pr(a_{gd}=1)=\prod_d (\sum_{k}\pi_kq_{kd})$ which is clearly different from $\sum_{k}(\pi_k\prod_d{q_{kd}})$. This explains why the hierarchical mixture model above can be used to describe the correlation among multiple studies. Since the mixture of $\mbox{\boldmath{$q_{k}$}}$s provides the key to model the cross-study correlation, each vector $\mbox{\boldmath{$q_{k}$}}$ is also called a ``correlation motif''.

A model with $K$ correlation motifs requires $O(KD)$ parameters in total. Usually, a small $K$ ($\ll 2^D$) is sufficient to capture the major correlation structure in the real data. Therefore, our method can be easily scaled up to deal with large $D$ scenarios. When $0<q_{kd}<1$, each $\mbox{\boldmath{$q_{k}$}}$ will be able to generate all $2^D$ configurations with non-zero probabilities. Thus, our model also retains the flexibility to allow all $2^D$ configurations of $\mbox{\boldmath{$a$}}_g$ to occur at individual gene level.

\subsection{Statistical Inference}
In reality, only $\mbox{\boldmath{$T$}}$ is observed. $\mbox{\boldmath{$\pi$}}$ and $\mbox{\boldmath{$Q$}}$ are unknown parameters. $\mbox{\boldmath{$A$}}$ and $\mbox{\boldmath{$B$}}$ are unobserved missing data. To infer the unknowns from $\mbox{\boldmath{$T$}}$, we first assume that $K$ is given and introduce a Dirichlet prior $Dir(2,...,2)$ for $\mbox{\boldmath{$\pi$}}$ and a Beta prior $B(2,2)$ for $q_{kd}$ such that:
\begin{eqnarray}
Pr(\mbox{\boldmath{$\pi$}},\mbox{\boldmath{$Q$}},\mbox{\boldmath{$A$}},\mbox{\boldmath{$B$}}|\mbox{\boldmath{$T$}})&\propto& \prod_{g=1}^{G}\prod_{k=1}^{K}\{\pi_k\prod_{d=1}^D[q_{kd}f_{d1}(t_{gd})]^{a_{gd}}[(1-q_{kd})f_{d0}(t_{gd})]^{1-a_{gd}}\}^{\delta{(b_g=k)}}\nonumber\\
&*&\prod_{k=1}^K\pi_k\prod_{k=1}^K\prod_{d=1}^Dq_{kd}(1-q_{kd})
\end{eqnarray}
Based on the above posterior distribution, an expectation-maximization (EM) algorithm can be derived to search for the posterior mode of $\mbox{\boldmath{$\pi$}}$ and $\mbox{\boldmath{$Q$}}$ (\citealp{rGelman}).
We chose the Dirichlet distribution $Dir(2,...,2)$ instead of $Dir(1,...,1)$ as prior since the mode of a Dirichlet distribution $Dir(\alpha_1,...,\alpha_K)$ for the $m^{th}$ component is $(\alpha_m-1)/(\sum_{k=1}^K\alpha_k-K)$, which is zero when $\alpha_m=1$ and not defined when all $\alpha_k$s are equal to one. As a result, in the EM iterations, when a motif is associated with very few genes such that $\sum_{g=1}^G E(\delta(b_g=m) | \mbox{\boldmath{$T$}}, \mbox{\boldmath{$\hat{\pi}$}}, \mbox{\boldmath{$\hat{Q}$}})$ is close to zero, the estimate of $\pi_m$ will become close to zero if we use $Dir(1,...,1)$. This will make the algorithm numerically unstable since the EM is implemented at logarithm scale (i.e., $log(\pi_m)$ instead of $\pi_m$ is used in the implementation to avoid underflow when multiplying multiple probabilities). The same reason explains why $B(2,2)$ was chosen as the prior for $q_{kd}$.

Using the estimated $\mbox{\boldmath{$\hat{\pi}$}}$ and $\mbox{\boldmath{$\hat{Q}$}}$, one can then compute $E(a_{gd} | \mbox{\boldmath{$T$}}, \mbox{\boldmath{$\hat{\pi}$}}, \mbox{\boldmath{$\hat{Q}$}}) = Pr(a_{gd} = 1| \mbox{\boldmath{$T$}}, \mbox{\boldmath{$\hat{\pi}$}}, \mbox{\boldmath{$\hat{Q}$}})$, the posterior probability that gene $g$ is differentially expressed in study $d$.
Next, we rank order genes in each study separately using $Pr(a_{gd} = 1| \mbox{\boldmath{$T$}}, \mbox{\boldmath{$\hat{\pi}$}}, \mbox{\boldmath{$\hat{Q}$}})$.
The ranked lists can be used to choose follow-up targets. Users can also provide a posterior probability cutoff to dichotomize genes into {\em differential\/} or {\em non-differential\/} genes in each study. The default cutoff is 0.5.

In order to choose the motif number $K$, we use Bayesian Information Criterion (BIC). Details of the EM algorithm and BIC computation are
provided in the Supplementary Materials A.1 and A.2.

{\em CorMotif\/} improves the differential expression detection by integrating information both across studies and across genes. $Pr(a_{gd} = 1| \mbox{\boldmath{$T$}}, \mbox{\boldmath{$\hat{\pi}$}}, \mbox{\boldmath{$\hat{Q}$}})$ can be decomposed as $\sum_{k=1}^K Pr(a_{gd} = 1| \mbox{\boldmath{$T$}}, \mbox{\boldmath{$\hat{\pi}$}}, \mbox{\boldmath{$\hat{Q}$}}, b_g=k)*Pr(b_g = k| \mbox{\boldmath{$T$}}, \mbox{\boldmath{$\hat{\pi}$}}, \mbox{\boldmath{$\hat{Q}$}})$. Here, $Pr(b_g = k| \mbox{\boldmath{$T$}}, \mbox{\boldmath{$\hat{\pi}$}}, \mbox{\boldmath{$\hat{Q}$}})$ is determined by jointly evaluating gene $g$'s expression data in all studies, and $Pr(a_{gd} = 1| \mbox{\boldmath{$T$}}, \mbox{\boldmath{$\hat{\pi}$}}, \mbox{\boldmath{$\hat{Q}$}}, b_g=k)$ contains information specific to study $d$. According to Bayes' theorem,  $Pr(a_{gd} = 1 | \mbox{\boldmath{$T$}}, \mbox{\boldmath{$\hat{\pi}$}}, \mbox{\boldmath{$\hat{Q}$}}, b_g=k) \propto Pr(t_{gd} | a_{gd} = 1, \mbox{\boldmath{$\hat{Q}$}}, b_g=k) \times Pr(a_{gd} = 1 | \mbox{\boldmath{$\hat{\pi}$}}, \mbox{\boldmath{$\hat{Q}$}}, b_g=k)$. $t_{gd}$ in the first term contains expression information for a given gene $g$ in study $d$. To compute its denominator, the limma approach also utilized information across genes to help with estimating the variance. Meanwhile, the second term $Pr(a_{gd} = 1 | \mbox{\boldmath{$\hat{\pi}$}}, \mbox{\boldmath{$\hat{Q}$}}, b_g=k)$  involves prior probabilities given by the correlation motifs (i.e., $\mbox{\boldmath{$\hat{q}_k$}}$s) which are estimated by examining data from all genes. Owing to this two-way information pooling (i.e., across both studies and genes), {\em CorMotif\/} uses information more effectively than methods based on only a single gene or a
single study. This is especially useful for analyzing studies with relatively weak signal-to-noise ratio.

\section{Simulations}
\subsection{Compared Methods}
\label{sec:Compared Method}
We compared {\em CorMotif \/} with six other methods: {\em separate
limma\/}, {\em  all concord\/}, {\em full motif\/}, {\em SAM\/}, {\em eb1\/}, {\em eb10best\/}. We did not compare the method in \cite{rJen} as no software was available for this method.
The {\em separate limma\/} approach analyzes each study separately using limma. The moderated t-statistics in each study are assumed to be a mixture of $t_{n_{0d}+n_d-2}$ and $(1+w_{d}/v_d)^{1/2}t_{n_{0d}+n_d-2}$. To better evaluate the gain from data integration, we matched this analysis to {\em CorMotif\/} as much as possible by running an EM algorithm similar to {\em CorMotif\/}  to compute the posterior probability for differential expression using 0.5 as default cutoff. Conceptually, this makes {\em separate
limma\/} equivalent to {\em CorMotif\/} with a single cluster ($K=1$), and the analysis produces the same gene ranking as limma in each study. {\em All concord\/} assumes that a gene is either
differentially expressed in all studies or non-differential in all studies (i.e., $\mbox{\boldmath{$a$}}_g = [1,1,\ldots,1]$ or $[0,0,\ldots,0]$). Conditional on $\mbox{\boldmath{$a$}}_g$, the model for $t_{gd}$ remains the same as {\em CorMotif\/} and limma.
{\em Full motif\/} assumes that genes fall into $2^D$ classes, corresponding to the $2^D$ possible $\mbox{\boldmath{$a$}}_g$ configurations. It can be viewed as a saturated version of the {\em CorMotif\/} model. All the other methods are applied to $x_{gdlj}$s directly. {\em SAM\/} (\citealp{rsam}) processes each study separately, whereas {\em eb1\/} and {\em eb10best\/} analyze all studies jointly.
The {\em eb1\/} method corresponds to the R package EBarrays with
lognormal-normal (LNN) and one cluster assumption (\citealp{reb1}).
The {\em eb10best\/} method is EBarrays with lognormal-normal and
multiple cluster assumption, and the cluster number is chosen as the one with the lowest AIC among 1 to
10 (\citealp{reb10}). We also tried XDE (\citealp{rXDE}). However, it took
extremely long computing time, usually 24 hours on a machine with 2.7GHz CPU and 4Gb RAM for 1000 iterations, for an analysis involving four studies. Moreover, 1000 iterations usually were not enough for XDE to converge for an analysis consisting of four studies, which was the smallest data we analyzed here. Therefore, XDE will not be compared hereinafter. {\em eb10best\/} failed to work when it was used to jointly analyze $\geq 7$ studies. {\em Full motif\/} and {\em eb1\/} failed when a dataset was composed of 20 studies.

\subsection{\label{subsec:ModelAssumpSimulation} Model-based Simulations}
We first tested {\em CorMotif\/} using simulations. In simulation 1, we generated 10,000 genes and four studies according to the four
differential patterns in Figure 2(a,b): 100 genes were differentially expressed in all four studies ($\mbox{\boldmath{$a$}}_g = [1,1,1,1]$);
400 genes were differential only in studies 1 and 2 ($[1,1,0,0]$); 400 genes were differential only in studies 2 and 3 ($[0,1,1,0]$); 9100
genes were non-differential ($[0,0,0,0]$). Each study had six samples: three cases and three controls. The variances $\sigma_{gd}^2$s were
simulated from a scaled inverse chi-square distribution $n_{0d}s_{0d}^2/\chi^2(n_{0d})$, where $n_{0d}=4$ and $s_{0d}^2=0.02$.
Given $\sigma_{gd}^2$, the expression values were generated using $x_{gdlj} \sim N(0,\sigma_{gd}^2$). Whenever $a_{gd}=1$, we drew $\mu_{gd}$ from $N(0,w_{0d}*\sigma_{gd}^2)$ where $w_{0d}=4$, and $\mu_{gd}$ was then added to the expression values of the three cases (i.e., $x_{gd1j}$s).

{\em CorMotif\/} was fit with the motif number $K$ varying from 1 to 10. The $K$ with the lowest BIC was chosen as the final motif number.
In this way, four motifs were reported, and they were very similar to the true underlying differential patterns (Figure \ref{Figure_simu} (c)).
To examine if {\em CorMotif\/} can improve gene ranking, for each study $d$ we counted the number of true differential genes (true positives),
$TP_d(r)$, among the top $r$ ranked genes for each method, and we plotted $TP_d(r)$ versus $r$ in Figure \ref{Figure_simu} (q,r,s,t).
{\em CorMotif\/} consistently performed among the best in all studies.
For instance, {\em CorMotif \/} identified 361 true differential genes among its top 500
gene list in study 1 (Figure \ref{Figure_simu}(q)). This performance was almost the same as the saturated model {\em full motif\/} which identified 362 true positives among the top 500 genes.
Among the other methods, {\em eb10best\/} identified 341, {\em all concord \/} identified 292, and the others identified fewer than 292
true positives among the top 500 genes. Thus, {\em CorMotif\/} detected at least 23.6\% more true positives compared to any other method
except {\em full motif\/} and {\em eb10best\/}. Both {\em full motif\/} and {\em eb10best\/} have the problem of exponentially growing parameter space
and will break down when the study number $D$ is large. In addition, {\em eb10best\/}
only identified 360 true positives among the top 1000 genes, whereas {\em CorMotif \/} identified 419, representing a 16.4\% improvement.

In {\em CorMotif\/}, we labeled genes as differential if the posterior probability $Pr(a_{gd} = 1| \mbox{\boldmath{$T$}}, \mbox{\boldmath{$\hat{\pi}$}}, \mbox{\boldmath{$\hat{Q}$}})>0.5$. Similarly, for {\em separate limma\/}, {\em all concord\/}, {\em full motif\/}, {\em eb1\/} and {\em eb10best\/}, differential expression was determined using their default posterior probability cutoff 0.5. For {\em SAM\/}, q-value cutoff 0.1 was used to call differential expression. At this cutoff, {\em SAM\/} reported similar number of genes with $\mbox{\boldmath{$a$}}_g = [0,0,0,0]$ (i.e., non-differential in all studies) compared with {\em CorMotif}. This  allowed us to meaningfully compare {\em SAM\/} and {\em CorMotif\/} in terms of their ability to find differential genes.
The confusion matrix in Table \ref{pdccormotif} shows that {\em CorMotif\/} was better at characterizing genes' true differential configurations compared to most other methods.
For instance, among the 400 $[0,1,1,0]$, 400 $[1,1,0,0]$ and 100 $[1,1,1,1]$ genes, {\em CorMotif \/} correctly reported differential label $a_{gd}$ in all four studies for 168, 151 and 33 genes
respectively. In contrast, {\em separate limma\/} only unmistakenly labeled 68, 57 and 4 genes respectively. {\em All concord\/} requires genes to have the same differential status in all studies.
As such, it lacks the flexibility to handle study-specific differential expression. It correctly identified 80 out of 100 $[1,1,1,1]$ genes, but none of the $[0,1,1,0]$ and $[1,1,0,0]$ genes
were correctly labeled as study-specific.
With the default cutoff, {\em eb1\/} and {\em eb10best\/} only labeled 62 and 0 out of 9100 $[0,0,0,0]$ genes as completely non-differential, compared to 9072 labeled by {\em CorMotif \/}.
In other words, {\em eb1\/} and {\em eb10best\/} reported more false positive differential expression events. At the same time, fewer $[0,1,1,0]$ and $[1,1,0,0]$ genes were correctly identified
by {\em eb1\/} (30 and 12 vs. 168 and 151 by {\em CorMotif \/}). Similarly, {\em SAM\/} was also poor at identifying the differential expression patterns $[1,1,1,1]$, $[1,1,0,0]$ and $[0,1,1,0]$. Among all the methods, only {\em full motif\/} performed slightly better than {\em CorMotif\/}. Even so, {\em CorMotif\/}
was able to perform close to this saturated model.
Adding up the diagonal elements in the confusion matrix for each method, {\em CorMotif\/}
unmistakenly assigned $\mbox{\boldmath{$a$}}_g$ labels to 9424 genes, whereas this number was
9164 for {\em separate limma\/}, 9175 for {\em all concord\/}, 9434 for {\em full motif\/},
168 for {\em eb1\/}, 509 for {\em eb10best\/}, and 9129 for {\em SAM\/}.

Using a similar approach, we performed simulations 2-4 which involved different study numbers and differential expression patterns shown in Figure \ref{Figure_simu}(e-p).
The complete results are shown in Supplemental Figure A.1 and Tables A.1-A.3. The conclusions were similar to simulation 1.
In particular, simulation 4 had 20 studies. {\em full motif\/}, {\em eb1\/} and {\em eb10best\/} all failed to run on this data.

\subsection{Simulations Based on Real Data}
In real data, the distributions for $x_{gdlj}$s may deviate from our model assumptions. Therefore, we further evaluated {\em CorMotif\/}
using simulations that retained the real data noise structure. In simulation 5, 24 Human U133 Plus 2.0 Affymetrix microarray samples were downloaded from four GEO experiments. Each experiment corresponds to a different tissue and consists of six biological replicates (Supplemental Table A.4). After RMA normalization, replicate samples in each experiment were split into three ``cases'' and three ``controls''. We then spiked in differential signals by adding random $N(0,1)$ deviates to the three cases according to patterns shown in Figure A.2 (a-b). Data simulated in this way were able to keep the background characteristics in real data. Simulation 5 is similar to simulations 1 and 2. {\em CorMotif\/} again recovered the underlying differential patterns. It
showed comparable differential gene detection performance to {\em full motif\/} and outperformed the other methods (Supplemental Figure A.2 (e-h), Table A.5). In a similar fashion, we performed simulations 6 and 7 based on real data (Supplemental Methods A.3 and Table A.4). These two simulations have the same differential signal patterns as simulations 3 and 4, respectively. Here, the motifs reported by {\em CorMotif\/} differ slightly from the underlying truth, but all the major correlation patterns were captured by the reported motifs. Once again, {\em CorMotif\/} performed the best in terms of differential gene detection (Supplemental Figure A.2 (i-x), Tables A.6-A.7), and {\em eb1\/}, {\em eb10best\/} and {\em full motif\/} failed to run when the study number increased (when they failed, their results were not shown).

\subsection{Motifs Are Parsimonious Representation of True Correlation Structures}
\label{sec:discordance}
As we use probability vectors to serve as motifs, it is possible that multiple weak patterns can be merged into a single motif. For instance, two complementary patterns [1,1,0,0] and [0,0,1,1] each with $n$ genes can be absorbed into a single motif with $\mbox{\boldmath{$q_k$}}=(0.5,0.5,0.5,0.5)$ having $2n$ genes. To illustrate, we conducted simulations 8-10 which were composed of the same samples as in simulation 5 and various proportions of differential expression patterns (Supplemental Figure A.3). In simulation 9 (Figure A.3 (i-l)), the relative abundance of two complementary block motifs ([1,1,0,0] and [0,0,1,1]) was small compared to the concordance motif [1,1,1,1], and they were absorbed into a single motif. In simulations 5, 8 and 10 (Figure A.3 (a-h),(m-p)), the complementary block motifs were more abundant, and the program successfully identified them as separate motifs. In general, we observed that weaker patterns were more likely to be merged than patterns with abundant data support. In all cases, however, {\em CorMotif\/} still provided the best gene ranking results compared to other methods (Supplemental Figure A.4). Supplemental Figures A.3 and A.4 also show that the higher the proportions of study-specific motifs (e.g., [1,1,0,0] and [0,0,1,1]), the better {\em CorMotif\/} will perform compared to the concordance analysis (i.e., {\em all concord\/}) in terms of ranking genes in each study. Together, the analyses here demonstrate that the correlation motifs only represent a parsimonious representation of the correlation structure supported by the available data. One should not expect {\em CorMotif\/} to always recover all the true underlying clusters exactly. In spite of this, our simulations show that {\em CorMotif} can still effectively utilize the correlation among studies to improve differential gene detection.

\section{Application to the Sonic Hedgehog (Shh) Signaling Data Sets}
We used {\em CorMotif\/} to analyze the SHH data in Table 1. The normalized data are available for download as Supplementary Table A.9. Datasets 1 and 2 compare SMO mutant mice with wild type mice (wt) and PTCH1 mutant with wild type, respectively, in
the 8 somite stage of developing embryos. Dataset 3 compares PTCH1 mutant with wild type in 13 somite stage.
Datasets 4 and 5 compare SHH mutant with wild type in developing head and limb, respectively.
Datasets 6 and 7 study gene expression changes in two SHH-related tumors, medulloblastoma and basal cell carcinoma (BCC),
compared to normal samples (control). Dataset 8 compares SMO mutant with wild type in the 13 somite stage of developing embryos.
{\em CorMotif\/} was applied to datasets 1-7. Dataset 8 was reserved for testing.

Five motifs were discovered (Figure \ref{Figure_real}(a,b)). Motif 1 mainly represents background. Motif 2 contains genes that have high probability to be differential in all studies. Genes in motif 3 tend to be differential in most studies except for the two involving PTCH1 mutant (i.e., studies 2 and 3). Most genes in motif 4 are not differential in the two studies involving the SHH mutant (i.e., studies 4 and 5) but tend to be differential in all other studies. Motif 5 mainly represents genes with differential expression in tumors (i.e., studies 6 and 7) but not in embryonic development (i.e., studies 1-5). In general, looking at the columns in Figure \ref{Figure_real}(a), the two studies involving tumors (6,7) are more similar to each other compared to other studies. The two PTCH1 mutant studies (2,3) are also relatively similar, and the same trend holds true for the two SHH mutant studies (4,5).

In this real data analysis, no comprehensive truth is available for evaluating differential expression calls. Without comprehensive knowledge about the true differential expression states of all genes in all cell types, we can only perform a partial evaluation based on existing knowledge. In this regard, we used dataset 8 as a test. Similar to dataset 1, this dataset compares SMO mutant with wild type. One expects that differential genes in these two datasets should be largely similar. Therefore, we used the top 217 differentially expressed genes detected by {\em separate limma\/} (at the posterior probability cutoff 0.5) in dataset 8 as gold standard to evaluate the gene ranking performance of different methods in dataset 1. Figure \ref{Figure_real}(c) shows that {\em CorMotif\/} again performed similar to {\em full motif\/} and outperformed all other methods. {\em eb10best\/} failed to run here. We note that since dataset 8 and datasets 2-7 represent more different biological contexts, one cannot use it as gold standard for evaluating these other datasets.

Finally, we examined well-studied SHH responsive target genes. Gli1, Ptch1, Ptch2, Hhip and Rab34 are known to be regulated by SHH signaling in somites and developing limb (\citealp{rVokesDevelop,rVokesGenes}).
Therefore, we expect these genes to be differential in studies 1, 2, 3 and 5. Figure \ref{Figure_real}(d) shows that {\em CorMotif\/}, {\em all concord\/} and {\em full motif\/} were able to correctly identify differential expression of these genes in all these studies, whereas {\em separate limma\/}, {\em SAM\/} and {\em eb1\/} failed to do so (they missed some cases). Table A.8 also shows that in many studies, {\em CorMotif\/},  {\em all concord\/} and  {\em full motif\/} provided better rank for these genes compared to {\em separate limma\/}, {\em SAM\/} and {\em eb1\/}. Hand2 is known to be a target of SHH signaling in developing limb but not in somites (\citealp{rVokesGenes}). While {\em separate limma\/}, {\em CorMotif\/}, {\em full motif\/} and {\em SAM \/} can correctly identify this, {\em all concord\/}  and {\em eb1\/} failed to do so. For {\em all concord\/}, since Hand2 was not differential in studies 1-4, 6 and 7, the method thinks that this gene is not differential in any study. Similarly, Hoxd13 is a limb specific target of SHH signaling \citep{rVokesGenes}. While the other methods correctly identified this, {\em all concord\/} failed again by claiming it to be differential in all studies. In all the genes examined, only {\em CorMotif\/} and {\em full motif\/} were able to correctly identify all known differential states.
Together, our analyses show that {\em CorMotif\/} offers unique advantage over the other methods in the integrative analysis of multiple gene expression studies.

\section{Discussion}
\label{sec4}
In summary, we have proposed a flexible and scalable approach for integrative analysis of differential gene expression in multiple studies. Using a few probability vectors instead of $2^D$ dichotomous vectors to characterize the differential expression patterns provides the key to circumvent the challenge of exponential growth of parameter space as the study number increases. The probabilistic nature of the motifs also allows all $2^D$ differential patterns to occur in the data at indiviual gene level.

The motif matrix $\mbox{\boldmath{$Q$}}$ can be viewed in two different ways. On one hand, each row of $\mbox{\boldmath{$Q$}}$ represents a cluster of genes with similar differential expression patterns across studies. Having many different motifs in $\mbox{\boldmath{$Q$}}$ is an indication that a concordance model, such as {\em all concord\/}, may not be sufficient to describe the correlation structure in the data.
On the other hand, each column of $\mbox{\boldmath{$Q$}}$ represents differential expression propensities of different gene classes in a given study. If two columns are similar, the corresponding studies share similar differential expression profiles (e.g., studies 6 and 7 in the SHH data are more similar to each other compared to the other studies in the same data).

{\em CorMotif\/} is computationally efficient. It took $\sim 0.5$ hour to analyze the SHH data for a given $K$, and $5.19$ hours in total to run all $K$s from 1 to 10. As a comparison, both {\em eb10best\/} and XDE failed, and {\em eb1} took 2.51 hours. {\em separate limma\/} (2.09 minutes) and {\em SAM\/} (1.71 minutes) were faster since each single study was processed separately each time. The relative efficiency of {\em CorMotif\/} is partly because we simplified the computation by modeling the moderated t-statistics $t_{gd}$ instead of the raw expression values $x_{gdlj}$s. In addition, we used EM instead of the more time-consuming MCMC to fit the model. Despite these simplifications, our results show that the present model robustly performs comparable or better than the alternative methods. A potential future work is to couple the correlation motif idea with more sophisticated models for the raw data $x_{gdlj}$ and explore whether the analysis can be improved further.

The {\em correlation motif \/} framework is general. Conceptually, one can modify the data generating distributions $f_{d0}$ and $f_{d1}$ to accommodate other data types, and use the same framework for a variety of meta-analysis problems. For example, with appropriate modification to $f_{d0}$s and $f_{d1}$s, the {\em correlation motif \/} idea should be directly applicable to RNA-seq data. Nevertheless, a systematic treatment of RNA-seq analysis is beyond the scope of this paper.

\section{Software}
\label{sec:software}

{\em CorMotif\/} is freely available as an R package in Bioconductor:

\noindent\url{http://www.bioconductor.org/packages/release/bioc/html/Cormotif.html}.

\section{Supplementary Material}
\label{sec:supplemental material}

Supplementary material is available online at
\href{http://biostatistics.oxfordjournals.org}%
{http://biostatistics.oxfordjournals.org}.

\section*{Acknowledgments}
The authors thank Drs. Andrew McMahon, Toyoaki Tenzen and Junhao Mao for providing the compiled SHH data, and
Robert B. Scharpf for his help with running XDE. The research is supported by the National Institutes of Health grant
R01HG006282.

{\it Conflict of Interest}: None declared.

\bibliographystyle{biorefs}
\bibliography{cormotifrefs}

\newpage

\begin{table}
\label{tab:shhdesign}
\begin{center}
\scalebox{0.9}{
\begin{tabular}{cccccc}
\hline Study ID & Condition 1 (case)& Sample No. & Condition 2 (control)& Sample No. & Reference\\
\hline
$~~1$ & 8somites\_smo & 3 & 8somites\_wt & 3 &\cite{rSHHAndrew0}\\
$~~2$ & 8somites\_ptc & 3 & 8somites\_wt& 3 &\cite{rSHHAndrew0}\\
$~~3$ & 13somites\_ptc & 3& 13somites\_wt&3 &\cite{rSHHAndrew0}\\
$~~4$ & head\_shh &  3& head\_wt&3 &\cite{rSHHAndrew0}\\
$~~5$ & limb\_shh & 3&limb\_wt& 3&\cite{rSHHAndrew0}\\
$~~6$ & Medulloblastoma\_tumor &3 & Medulloblastoma\_control& 2 &\cite{rSHHMao} \\
$~~7$ & BCC\_tumor &3& BCC\_control & 3&\cite{rSHHMao}\\
$~~8$ & 13somites\_smo &3& 13somites\_wt& 3&\cite{rSHHAndrew0}\\
\hline
\end{tabular}}
\caption{SHH microarray data description. 8somites and 13somites indicate
two different developmental stages of embryos; smo indicates mice with mutant Smo;
ptc stands for mice with mutant Ptch1; wt means wild type; shh represents Shh mutant.
Medulloblastoma and BCC (basal cell carcinoma) are two types of tumors.}
\end{center}
\end{table}

\begin{table}[htbp]
\begin{center}
\begin{tabular}{c|rrrrc}
\hline Method & Motif pattern & \multicolumn{1}{c}{$c(0,0,0,0)$} &
\multicolumn{1}{c}{$c(0,1,1,0)$} &
\multicolumn{1}{c}{$c(1,1,0,0)$} &\multicolumn{1}{c}{$c(1,1,1,1)$}\\
\hline
{\em CorMotif \/} &$~~c(0,0,0,0)$ & 9072 & 161 & 165 & 16 \\
&$~~c(0,1,1,0)$ &  3 & 168 & 3 & 7 \\
&$~~c(1,1,0,0)$ &  3 & 2 & 151 & 6 \\
&$~~c(1,1,1,1)$ &  0 & 1 & 0 & 33 \\
&$~~other$ &       22 & 68 & 81 & 38 \\
\hline
{\em separate limma\/} & $~~c(0,0,0,0)$ & 9035 & 144 & 144 & 16 \\
&$~~c(0,1,1,0)$ & 0 & 68 & 0 & 5 \\
&$~~c(1,1,0,0)$ &  0 & 0 & 57 & 6 \\
&$~~c(1,1,1,1)$ &  0 & 0 & 0 & 4 \\
&$~~other$ &   65 & 188 & 199 & 69 \\
\hline
{\em all concord\/} & $~~c(0,0,0,0)$ & 9095 & 236 & 236 & 20 \\
& $~~c(0,1,1,0)$ & 0 &    0 & 0   & 0\\
& $~~c(1,1,0,0)$ &  0 &    0 & 0   & 0\\
&$~~c(1,1,1,1)$ &  5 & 164 & 164 & 80 \\
&$~~other$ &      0 &   0 & 0  & 0\\
 \hline
{\em full motif \/}&$~~c(0,0,0,0)$ & 9072 & 161 & 164 & 16 \\
&$~~c(0,1,1,0)$ &  4 & 172 & 4 & 7 \\
&$~~c(1,1,0,0)$ &  3 & 2 & 155 & 6 \\
&$~~c(1,1,1,1)$ &  0 & 1 & 0 & 35 \\
&$~~other$ &      21 & 64 & 77 & 36 \\
\hline
{\em eb1\/} &$~~c(0,0,0,0)$ & 62 & 0 & 2 & 0 \\
&$~~c(0,1,1,0)$ & 2178 & 30 & 22 & 3 \\
&$~~c(1,1,0,0)$ & 569 & 7 & 12 & 0 \\
&$~~c(1,1,1,1)$ & 753 & 34 & 32 & 64 \\
&$~~others$ &      5538 & 329 & 332 & 33 \\
 \hline
{\em eb10best\/}&$~~c(0,0,0,0)$ & 0 & 0 & 0 & 1 \\
&$~~c(0,1,1,0)$ & 316 & 220 & 16 & 10 \\
&$~~c(1,1,0,0)$ & 180 & 23 & 226 & 10 \\
&$~~c(1,1,1,1)$ & 5789 & 77 & 52 & 63 \\
&$~~other$ &     2815 & 80 & 106 & 16 \\
 \hline
 {\em SAM\/}&$~~c(0,0,0,0)$ & 9099 & 256 & 279  & 48  \\
&$~~c(0,1,1,0)$ & 0  & 20  &  0  &  3 \\
&$~~c(1,1,0,0)$ &  0  &  0 &   9   & 2 \\
&$~~c(1,1,1,1)$ & 0  &  0&    0  &  1 \\
&$~~other$ &     1 & 124&  112  & 46 \\
\hline
\end{tabular}
\caption{Confusion matrix for simulation 1. The column labels indicate the
true underlying patterns and the row labels represent the reported configurations
at gene level. For {\em CorMotif\/}, {\em separate
limma\/}, {\em all concord\/}, {\em full motif\/}, {\em eb1\/} and {\em eb10best\/}, differential expression in each study is determined using their default posterior
probability cutoff 0.5. For {\em SAM\/}, q-value cutoff 0.1 was used to call differential expression. This yields similar number of correct classifications
for pattern $[0,0,0,0]$ compared with {\em CorMotif}. } \label{pdccormotif}
\end{center}
\end{table}

\begin{figure}[!b]
\begin{center}
\includegraphics[width=1.1\textwidth]{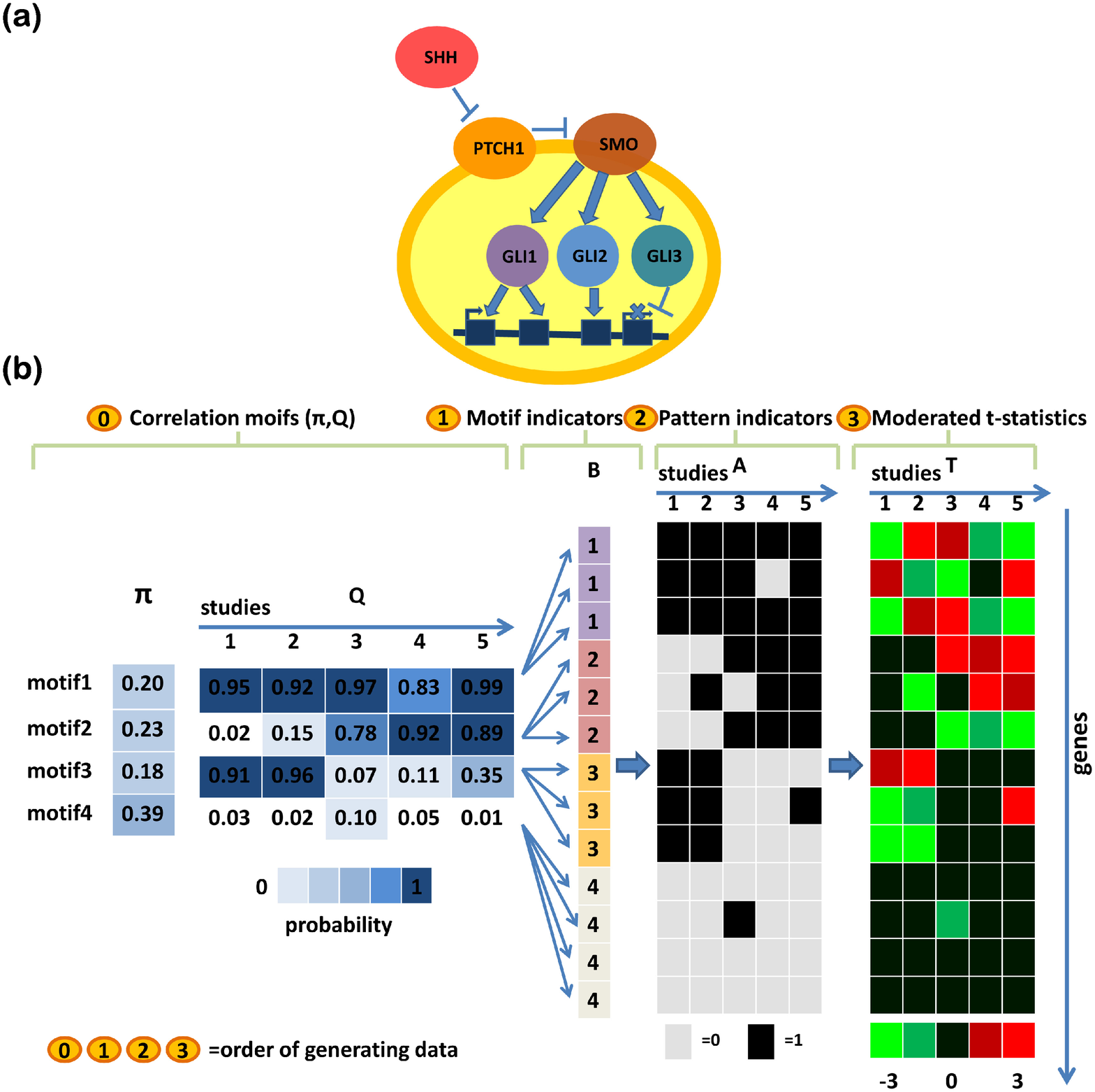}
\caption{(a) A cartoon illustration of SHH pathway. (b) A numerical example of the data generating model. There exist four motifs in the dataset, with the abundance $\mbox{\boldmath{$\pi$}}=(0.2,0.23,0.18,0.39)$. Each row of the $\mbox{\boldmath{$Q$}}$ matrix represents a motif and each column corresponds to a study. Thus, $\mbox{\boldmath{$q$}}_{kd}$ indicates the probability for genes belonging to motif $k$ to be differentially expressed in study $d$. For example, the probability for genes belonging to motif 1 to be differentially expressed in study 4 is 0.83. The gray scale of the cells in $\mbox{\boldmath{$\pi$}}$ and $\mbox{\boldmath{$Q$}}$ illustrates the probability value, with white indicating probability 0 and dark blue representing probability 1. Given $\mbox{\boldmath{$\pi$}}$ and $\mbox{\boldmath{$Q$}}$, each gene is assigned a motif indicator $b_g$. For instance, the fifth gene belongs to motif 2 (indicated by a cell of shallow red color with a number ``2''). Next, the configuration of the fifth gene, $[a_{51},a_{52},a_{53},a_{54},a_{55}]$, is generated according to $\mbox{\boldmath{$q$}}_2=(0.02,0.15,0.78,0.92,0.89)$. As a result, the fifth gene is differentially expressed in study 2,4 and 5. Finally, the moderated t-statistic $t_{5d}$ within each study $d$ is produced according to the configuration $a_{5d}$.}
\label{genproc}
\end{center}
\end{figure}

\begin{figure}
\begin{center}
\includegraphics[width=1.2\textwidth,angle=270]{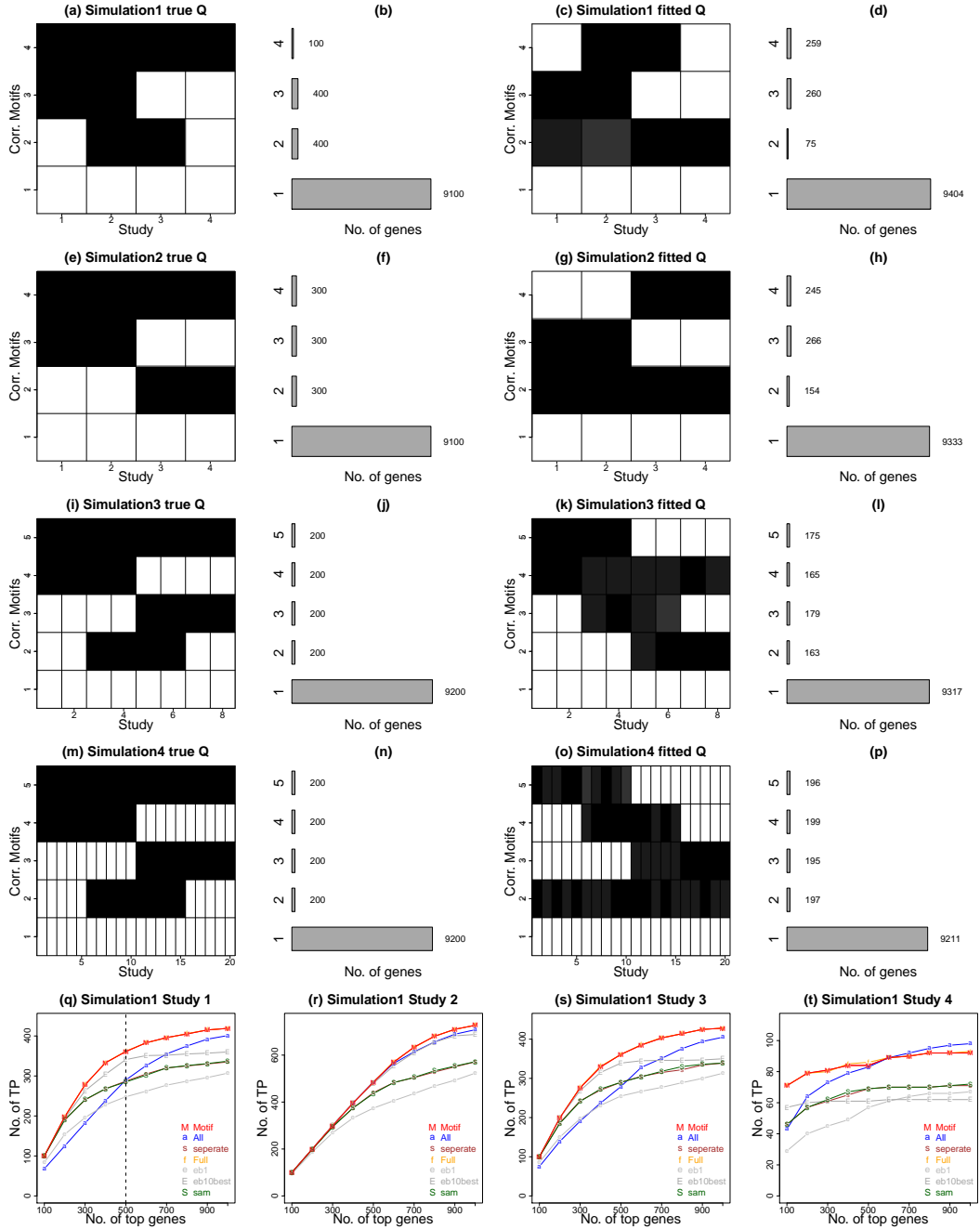}
\caption{Results for the model assumption based simulations. Also see Supplementary Figure A.1. (a),(e),(i),(m) Motif patterns for simulations 1-4. The $\mbox{\boldmath{$Q$}}$ of the true motifs in the simulated data. (b),(f),(j),(n) The true number of genes belonging to each motif in the simulated data (i.e., $\mbox{\boldmath{$\pi$}}*G$). (c),(g),(k),(o) The estimated $\mbox{\boldmath{$\hat{Q}$}}$ from the learned motifs. (d),(h),(l),(p) The estimated number of genes belonging to each learned motif (i.e., $\mbox{\boldmath{$\hat{\pi}$}}*G$). In
the $\mbox{\boldmath{$Q$}}$ pattern graphs in columns 1 and 3, each row indicates a motif
pattern and each column represents a study. The gray scale of the
cell $(k,d)$ demonstrates the probability of differential expression
in study $d$ for pattern $k$. Black means 1 and while means 0. Each row of the bar chart for ($\mbox{\boldmath{$\pi$}}*G$) corresponds
to the motif pattern in the same row of the $\mbox{\boldmath{$Q$}}$ pattern graph. It can be seen that motif patterns learned
by {\em CorMotif \/} are similar to the true underlying motif
patterns. (q)-(t) Gene ranking performance of different methods in simulation 1. $TP_d(r)$, the number of
genes that are truly differentially expressed in study $d$ among the
top $r$ ranked genes by a given method, is plotted against the rank cutoff $r$.}
\label{Figure_simu}
\end{center}
\end{figure}

\begin{figure}
\begin{center}
\includegraphics[width=\textwidth]{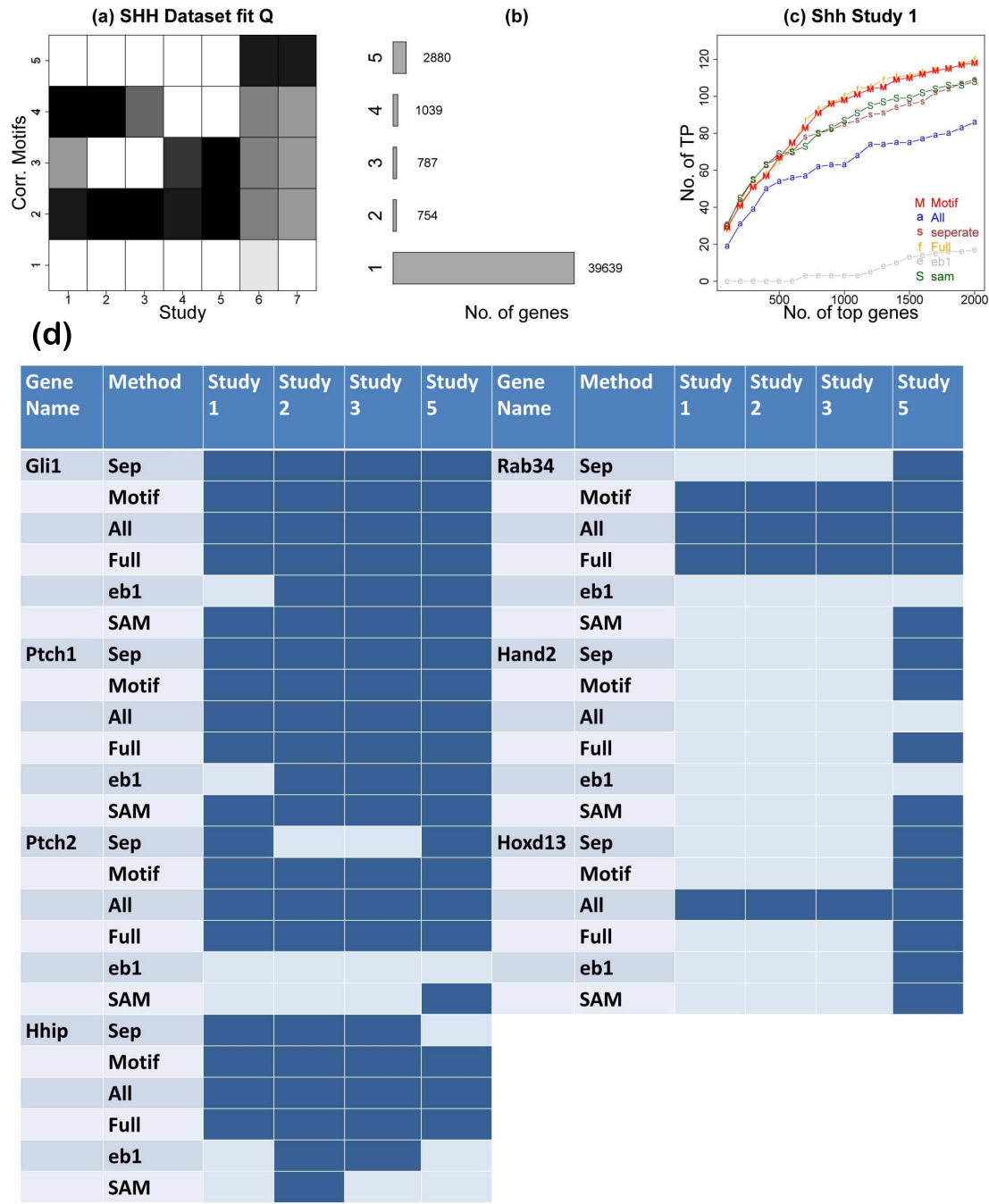}
\caption{Results for the SHH data. (a)-(b) Motif patterns learned from the SHH data composed of 7 studies. (c)
Gene ranking performance for SHH study 1. The genes differentially expressed in dataset 8 (13somites\_smo vs.
13somites\_wt) were obtained using {\em separate limma\/}. They were used as the gold standard. $TP_d(r)$, the number of
genes in dataset 1 that are truly differentially expressed among the
top $r$ ranked genes by each method, is plotted against the rank cutoff $r$. (d) Differential status claimed by each method for known SHH pathway genes. Dark blue indicates differential expression and light grey represents non-differential expression.}
\label{Figure_real}
\end{center}
\end{figure}

\end{document}


\title{Supplementary Materials to Joint Analysis of Differential Gene Expression in Multiple Studies using Correlation Motifs}

\author{YINGYING WEI, HONGKAI JI$^\ast$,\\
[4pt]
\textit{Department of Biostatistics, Johns Hopkins University Bloomberg School of Public Health, Baltimore, Maryland, USA}
\\[2pt]
{hji@jhsph.edu}}

\markboth%
{YY Wei and HK Ji}
{Correlation Motif Discovery Supplementary Materials}

\maketitle

\footnotetext{To whom correspondence should be addressed.}

\section{The em algorithm used in cormotif}
This section presents the EM algorithm used to search for posterior mode of $\mbox{\boldmath{$\hat{\pi}$}}$ and
$\mbox{\boldmath{$\hat{Q}$}}$ of the distribution $Pr(\mbox{\boldmath{$\pi$}},\mbox{\boldmath{$Q$}}|\mbox{\boldmath{$T$}})=\sum_{\mbox{\boldmath{$A$}},\mbox{\boldmath{$B$}}}Pr(\mbox{\boldmath{$\pi$}},\mbox{\boldmath{$Q$}},\mbox{\boldmath{$A$}},\mbox{\boldmath{$B$}}|\mbox{\boldmath{$T$}})$. In the EM algorithm, $\mbox{\boldmath{$A$}}$ and $\mbox{\boldmath{$B$}}$ are missing data. The algorithm iterates between the E-step and the M-step.

In the E-step, one evaluates the Q-function $Q(\mbox{\boldmath{$\pi$}},\mbox{\boldmath{$Q$}}|\mbox{\boldmath{$\hat{\pi}^{old}$}},\mbox{\boldmath{$\hat{Q}^{old}$}})$ which is defined as $E_{old}[\ln Pr(\mbox{\boldmath{$\pi$}},\mbox{\boldmath{$Q$}},\mbox{\boldmath{$A$}},\mbox{\boldmath{$B$}}|\mbox{\boldmath{$T$}})]$. Here the expectation is taken with
respect to distribution $Pr(\mbox{\boldmath{$A$}},\mbox{\boldmath{$B$}}|\mbox{\boldmath{$T$}},\mbox{\boldmath{$\hat{\pi}^{old}$}},\mbox{\boldmath{$\hat{Q}^{old}$}})$, abbreviated as $Pr_{old}(\mbox{\boldmath{$A$}},\mbox{\boldmath{$B$}})$, where $\mbox{\boldmath{$\hat{\pi}^{old}$}},\mbox{\boldmath{$\hat{Q}^{old}$}}$ are the parameter estimates obtained from the last iteration.

We have
\begin{eqnarray}
\lefteqn{\ln Pr(\mbox{\boldmath{$\pi$}},\mbox{\boldmath{$Q$}},\mbox{\boldmath{$A$}},\mbox{\boldmath{$B$}}|\mbox{\boldmath{$T$}})=\sum_{g=1}^G\sum_{k=1}^K\delta{(b_g=k)}\ln\pi_k}
\nonumber\\
&&+\sum_{g=1}^G\sum_{k=1}^K\delta(b_g=k) \left\{ \sum_{d=1}^D a_{gd}[\ln
q_{kd}+\ln f_{d1}(x_{gd})]+\sum_{d=1}^D(1-a_{gd})[\ln(1-q_{kd})+\ln
f_{d0}(x_{gd})] \right\}
\nonumber\\
&&+\sum_{k=1}^K\ln\pi_k+\sum_{k=1}^K\sum_{d=1}^D[\ln
q_{kd}+\ln(1-q_{kd})]+constant
\end{eqnarray}

Therefore,
\begin{eqnarray}
&&Q(\mbox{\boldmath{$\pi$}},\mbox{\boldmath{$Q$}}|\mbox{\boldmath{$\hat{\pi}^{old}$}},\mbox{\boldmath{$\hat{Q}^{old}$}})=E_{old}[\ln Pr(\mbox{\boldmath{$\pi$}},\mbox{\boldmath{$Q$}},\mbox{\boldmath{$A$}},\mbox{\boldmath{$B$}}|\mbox{\boldmath{$T$}})]
\nonumber\\
&&=\sum_{g=1}^G\sum_{k=1}^K\ln\pi_kE_{old}(\delta(b_g=k))
\nonumber\\
&&+\sum_{g=1}^G\sum_{k=1}^K\sum_{d=1}^D[\ln q_{kd}+\ln
f_{d1}(x_{gd})]E_{old}(\delta(b_g=k)a_{gd})
\nonumber\\
&&+\sum_{g=1}^G\sum_{k=1}^K\sum_{d=1}^D[\ln(1-q_{kd})+\ln
f_{d0}(x_{gd})]E_{old}(\delta(b_g=k)(1-a_{gd}))
\nonumber\\
&&+\sum_{k=1}^K\ln\pi_k+\sum_{k=1}^K\sum_{d=1}^D[\ln
q_{kd}+\ln(1-q_{kd})]+constant
\nonumber\\
\end{eqnarray}

In the M-step, one finds $\mbox{\boldmath{$\pi$}}$ and $\mbox{\boldmath{$Q$}}$ that maximize the Q-function $Q(\mbox{\boldmath{$\pi$}},\mbox{\boldmath{$Q$}}|\mbox{\boldmath{$\hat{\pi}^{old}$}},\mbox{\boldmath{$\hat{Q}^{old}$}})$. Denote them as $\mbox{\boldmath{$\hat{\pi}^{new}$}}$ and $\mbox{\boldmath{$\hat{Q}^{new}$}}$ and they will be used in next iteration.

By solving
\begin{equation}
\frac{\partial Q(\mbox{\boldmath{$\pi$}},\mbox{\boldmath{$Q$}}|\mbox{\boldmath{$\hat{\pi}^{old}$}},\mbox{\boldmath{$\hat{Q}^{old}$}})}{\partial\pi_k}=0
\end{equation}

\begin{equation}
\frac{\partial Q(\mbox{\boldmath{$\pi$}},\mbox{\boldmath{$Q$}}|\mbox{\boldmath{$\hat{\pi}^{old}$}},\mbox{\boldmath{$\hat{Q}^{old}$}})}{\partial
q_{kd}}=0
\end{equation}

We have
\begin{equation}
\hat{\pi}_k^{new}=\frac{\sum_{g=1}^GPr_{old}(b_g=k)+1}{G+K}
\end{equation}

\begin{equation}
\hat{q}_{kd}^{new}=\frac{\sum_{g=1}^GPr_{old}(b_g=k,a_{gd}=1)+1}{\sum_{g=1}^GPr_{old}(b_g=k)+2}
\end{equation}

In the formulae above, $Pr_{old}(b_g=k)$ and $Pr_{old}(b_g=k, a_{gd}=1)$ can be computed as below
\begin{equation}
Pr_{old}(b_g=k)=\frac{\hat{\pi}_k^{(old)}\prod_{d=1}^D[\hat{q}_{kd}^{(old)}f_{d1}(t_{gd})+(1-\hat{q}_{kd}^{(old)})f_{d0}(t_{gd})]}{\sum_{l=1}^K\hat{\pi}_l^{(old)}\prod_{d=1}^D[\hat{q}_{ld}^{(old)}f_{d1}(t_{gd})+(1-\hat{q}_{ld}^{(old)})f_{d0}(t_{gd})]}
\end{equation}

\begin{eqnarray}
&&Pr_{old}(b_g=k,a_{gd}=1)=Pr_{old}(a_{gd}=1|b_g=k)*Pr_{old}(b_g=k)\nonumber\\
&&=\frac{\hat{q}_{kd}^{(old)}f_{d1}(t_{gd})}{\hat{q}_{kd}^{(old)}f_{d1}(t_{gd})+(1-\hat{q}_{kd}^{(old)})f_{d0}(t_{gd})}
Pr_{old}(b_g=k) \nonumber\\
\end{eqnarray}

Therefore, we can iteratively use the EM algorithm to obtain the
estimates for $\pi$ and $Q$.

\section{Bayesian information criterion (bic) for choosing k}
BIC  is computed as
\begin{eqnarray}
BIC(K) &=&-2*\ln Pr(\mbox{\boldmath $T$}|\mbox{\boldmath $\pi$}, \mbox{\boldmath$Q$}) + (K-1+K*D)*\ln G
\\
&&=-2*\sum_{g=1}^G\ln \left[ \sum_{k=1}^K \{\pi_k \prod_{d=1}^D [q_{kd}f_{d1}(t_{gd})+(1-q_{kd})f_{d0}(t_{gd})] \} \right]+(K-1+K*D)*\ln G \nonumber\\\nonumber
\end{eqnarray}

BIC for different values of $K$ are calculated and the $K$ corresponding to the model that achieves the smallest BIC is chosen. Here $K$ is the number of motifs in the data and $K-1$ is the number of parameters for $\mbox{\boldmath $\pi$}$. $KD$ is the number of parameters involved in $\mbox{\boldmath $Q$}$. $G$ is the gene number.

\section{Data for real data based simulations}
Simulations 5-10 were based on real data characteristics. Each simulation contained multiple studies, and each study was composed of six samples from the same GEO experiment with the same biological condition as detailed in Table A.4. The six samples were further split into three pseudo cases and three pseudo controls. They were used as the simulated background since one does not expect differential signals between replicate samples. We then spiked in differential signals by adding random $N(0,1)$ numbers to the three cases according to the patterns shown in Figures A.2 (a-b,i-j,q-r) and A.3(a-b,e-f,i-j,m-n). Data simulated in this way were able to keep the background characteristics in real data.

\newpage
\begin{figure}
\begin{center}
\includegraphics[width=0.75\textwidth,angle=270]{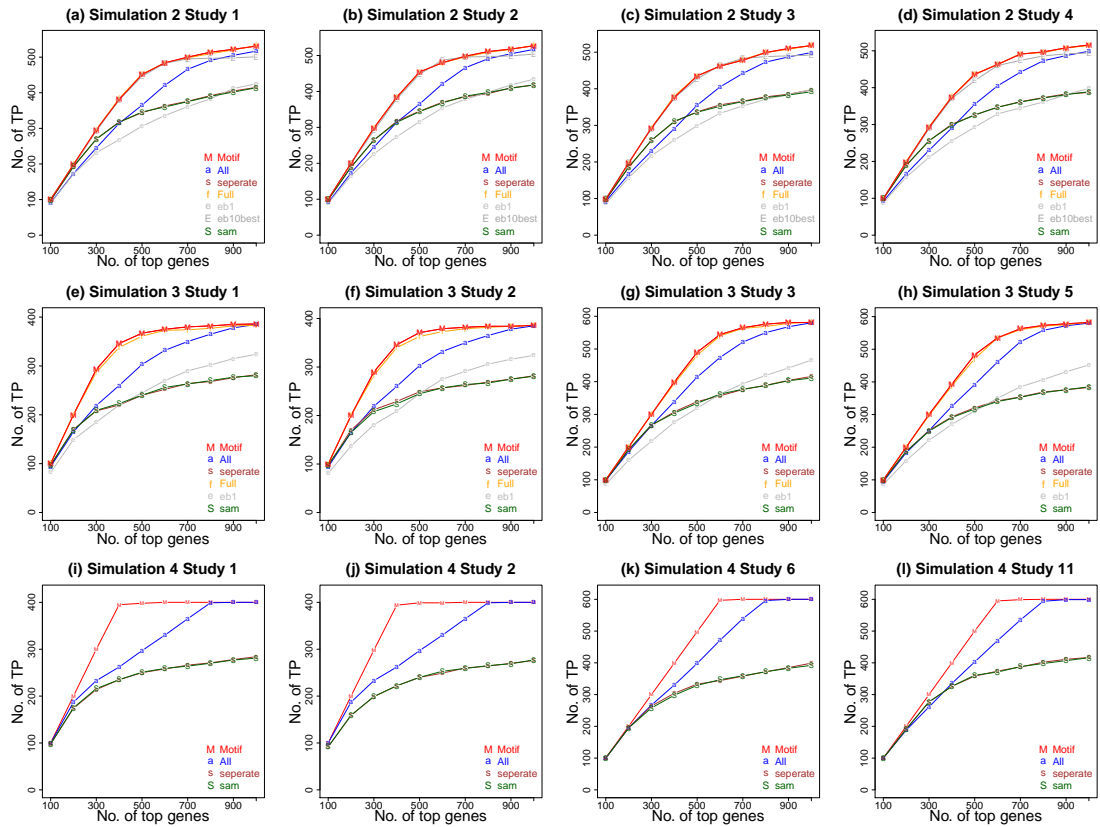}
\caption{Gene ranking performance for simulations 2, 3 and 4. $TP_d(r)$, the number of
genes that are truly differentially expressed in study $d$ among the
top $r$ ranked genes by a given method, is plotted against the rank cutoff $r$. Simulations 3 and 4 contain more than four studies, and results for four representative studies are shown. (a)-(d) Simulation 2. (e)-(h) Simulation 3. Studies 1 and 2 are representative for patterns in studies 1, 2 and 7, 8; studies 3 and 5 are representative for patterns in studies 3 to 6. (i)-(l) Simulation 4. Studies 1 and 2 are representative for patterns in studies 1-5 and 16-20; studies 6 and 11 are representative for patterns in studies 6-15.} \label{roc_Dataset1}
\end{center}
\end{figure}

\begin{figure}
\begin{center}
\includegraphics[width=1.1\textwidth,angle=270]{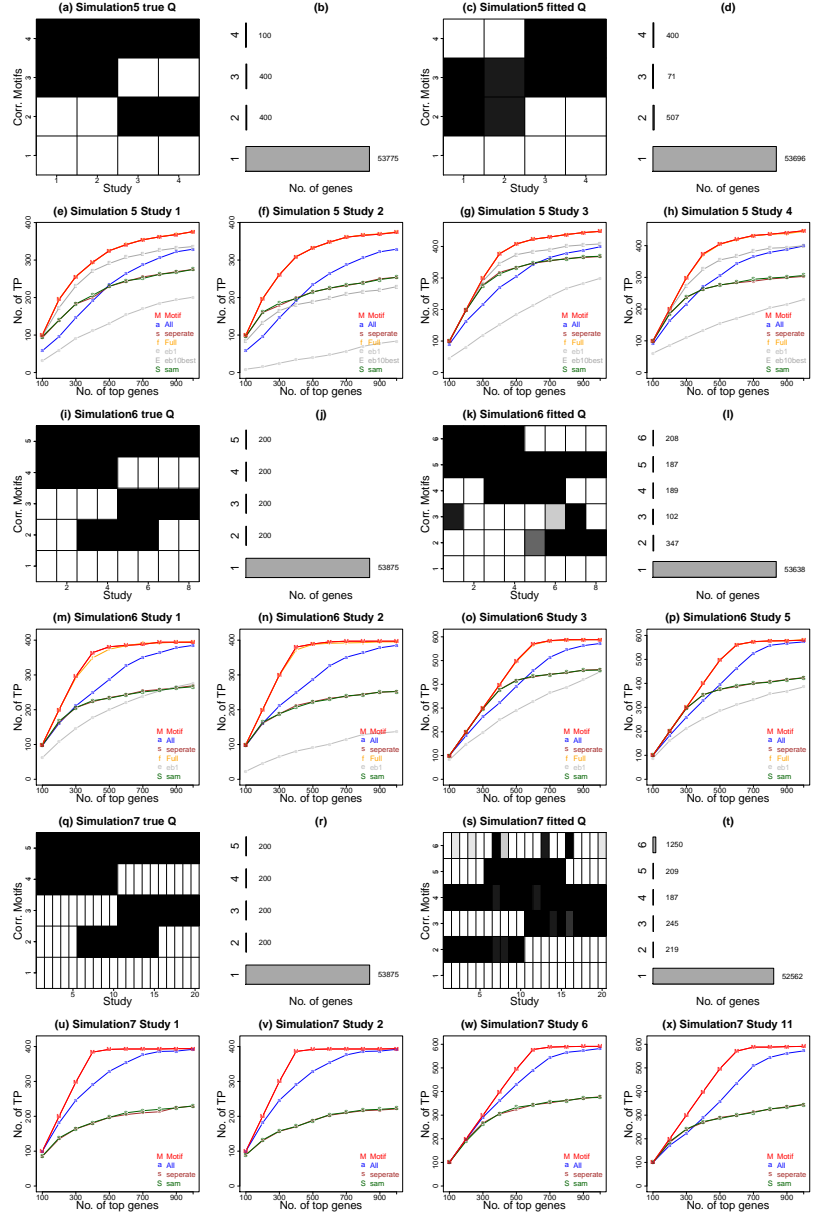}
\caption{Motif patterns and gene ranking performance for simulations 5, 6 and 7. $TP_d(r)$, the number of
genes that are truly differentially expressed in study $d$ among the
top $r$ ranked genes by given method, is plotted against the rank cutoff $r$. (a)-(d) True and estimated motif patterns for simulation 5. (e)-(h) Gene ranking performance for simulation 5. (i-l) Motif patterns for simulation 6. (m)-(p) Gene ranking performance for simulation 6. Studies 1 and 2 are representative for patterns in studies 1, 2 and 7, 8; studies 3 and 5 are representative for patterns in studies 3 to 6. (q-t) Motif patterns for simulation 7. (u)-(x) Gene ranking performance for simulation 7. Studies 1 and 2 are representative for patterns in studies 1-5 and 16-20; studies 6 and 11 are representative for patterns in studies 6-15.} \label{roc_Dataset1}
\end{center}
\end{figure}

\begin{figure}
\begin{center}
\includegraphics[width=0.9\textwidth,angle=270]{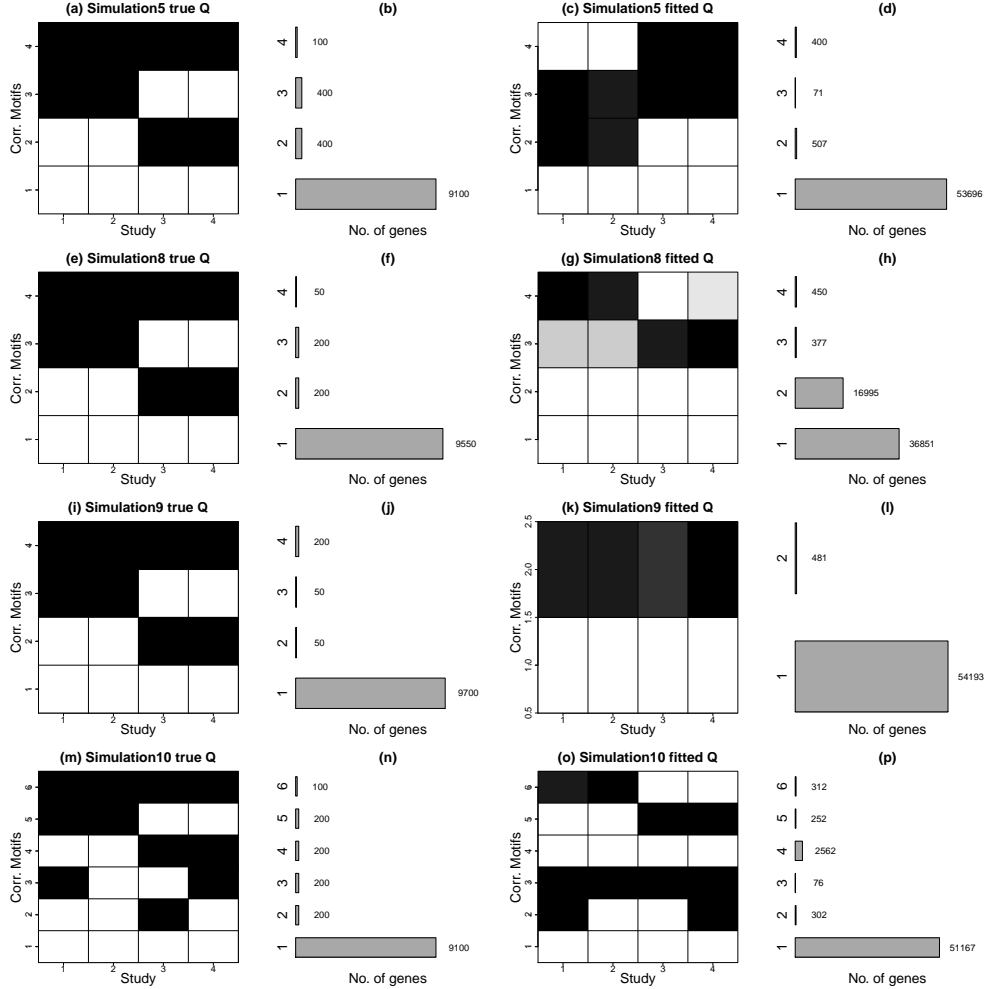}
\caption{Motif patterns for simulations 5, 8, 9 and 10. (a),(e),(i),(m) The $\mbox{\boldmath{$Q$}}$ for the true underlying motifs in the simulated data. (b),(f),(j),(n) The true number of genes belonging to each motif in the simulated data (i.e., $\mbox{\boldmath{$\pi$}}*G$). (c),(g),(k),(o) The estimated $\mbox{\boldmath{$\hat{Q}$}}$ for the learned motifs. (d),(h),(l),(p) The estimated number of genes belonging to each learned motif (i.e., $\mbox{\boldmath{$\hat{\pi}$}}*G$). In
the $\mbox{\boldmath{$Q$}}$ pattern graph (columns 1 and 3), each row indicates a motif
pattern and each column represents a study. The gray scale of the
cell $(k,d)$ demonstrates the probability of differential expression
in study $d$ for pattern $k$. Each row of the bar chart for ($\mbox{\boldmath{$\pi$}}*G$) corresponds
to the motif pattern in the same row of the $\mbox{\boldmath{$Q$}}$ graph. The motif patterns learned
by {\em CorMotif\/} are similar to the true underlying motif
patterns. It can be seen that complementary block motifs, such as [1,1,0,0] and [0,0,1,1], are not likely to
be absorbed into merged motifs if their relative proportions are not low.}
\end{center}
\end{figure}

\begin{figure}
\begin{center}
\includegraphics[width=0.9\textwidth,angle=270]{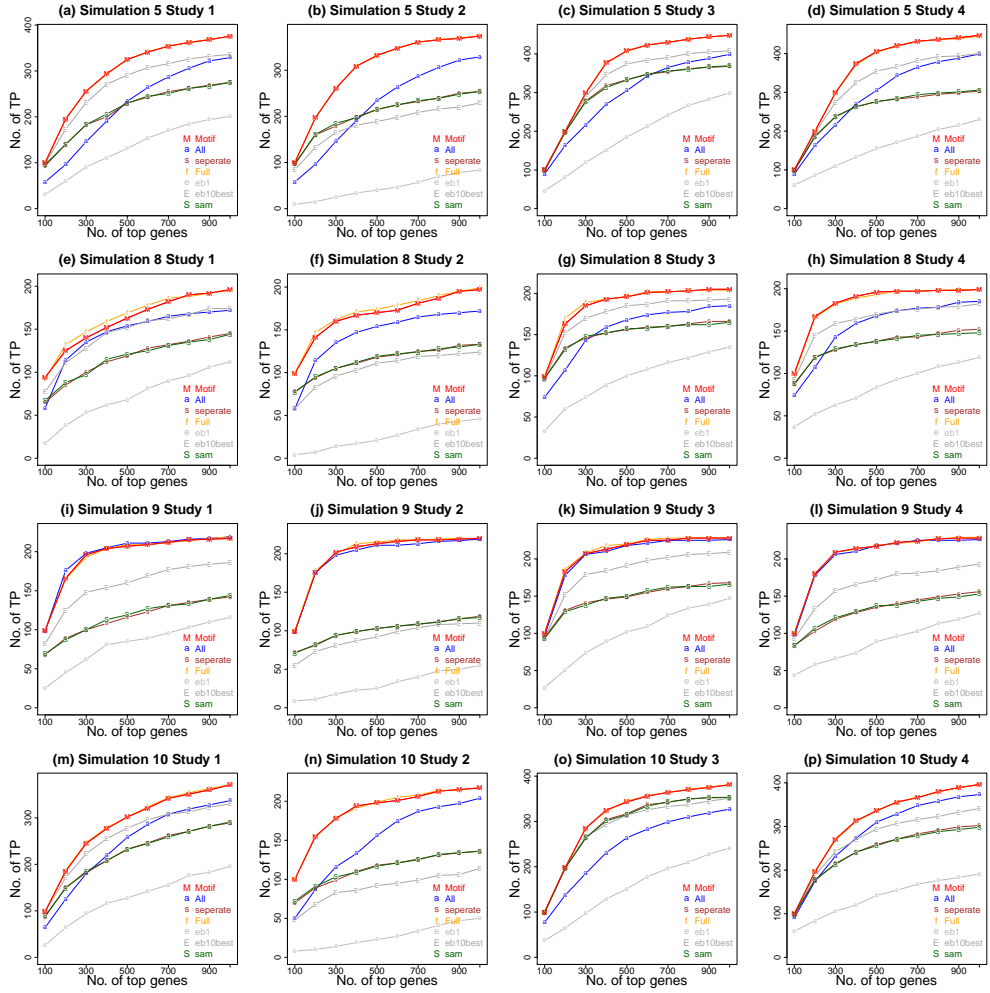}
\caption{Gene ranking performance for simulations 5, 8, 9 and 10. $TP_d(r)$, the number of
genes that are truly differentially expressed in study $d$ among the
top $r$ ranked genes by a given method, is plotted against the rank cutoff $r$. (a)-(d) Simulation 5. (e)-(h) Simulation 8. (i)-(l) Simulation 9. (m)-(p) Simulation 10.} \label{roc_concordance}
\end{center}
\end{figure}

\begin{table}[H]
\caption{Confusion matrix for simulation 2.  The column labels indicate the
true underlying patterns and the row labels represent the learned configurations.} \label{pdccormotif2}
\begin{center}
\scalebox{0.8}{
\begin{tabular}{c|rrrrc}
\hline Method & Motif pattern & \multicolumn{1}{c}{$c(0,0,0,0)$} &
\multicolumn{1}{c}{$c(0,0,1,1)$} &
\multicolumn{1}{c}{$c(1,1,0,0)$} &\multicolumn{1}{c}{$c(1,1,1,1)$}\\
\hline
{\em Cormotif\/} &$~~c(0,0,0,0)$ & 9069 & 122 & 99 & 54\\
&$~~c(0,0,1,1)$ & 7 & 127 & 0 & 30 \\
&$~~c(1,1,0,0)$ &   3 & 0 & 153 & 29 \\
&$~~c(1,1,1,1)$ &  0 & 1 & 1 & 89 \\
&$~~other$ &      21 & 50 & 47 & 98 \\
\hline
{\em separate limma\/} & $~~c(0,0,0,0)$ & 9024 & 112 & 89 & 58 \\
&$~~c(0,0,1,1)$ & 1 & 44 & 0 & 13 \\
&$~~c(1,1,0,0)$ &   0 & 0 & 57 & 17 \\
&$~~c(1,1,1,1)$ &  0 & 0 & 0 & 8 \\
&$~~other$ &     75 & 144 & 154 & 204 \\
\hline
{\em all concord\/} & $~~c(0,0,0,0)$ & 9094 & 180 & 166 & 76 \\
& $~~c(0,0,1,1)$ & 0 &    0 & 0   & 0\\
& $~~c(1,1,0,0)$ &  0 &    0 & 0   & 0\\
&$~~c(1,1,1,1)$ &  6 & 120 & 134 & 224 \\
&$~~other$ &      0 &   0 & 0  & 0\\
 \hline
{\em full motif \/}&$~~c(0,0,0,0)$ & 9069 & 122 & 99 & 54 \\
&$~~c(0,0,1,1)$ & 7 & 130 & 0 & 33 \\
&$~~c(1,1,0,0)$ &  5 & 0 & 160 & 29 \\
&$~~c(1,1,1,1)$ &  0 & 1 & 1 & 99 \\
&$~~other$ &      19 & 47 & 40 & 85 \\
\hline
{\em eb1\/} &$~~c(0,0,0,0)$ & 4693 & 20 & 8 & 5 \\
&$~~c(0,0,1,1)$ &  376 & 65 & 1 & 8 \\
&$~~c(1,1,0,0)$ &  474 & 1 & 74 & 10 \\
&$~~c(1,1,1,1)$ & 365 & 131 & 132 & 238 \\
&$~~other$ &     3192 & 83 & 85 & 39 \\
 \hline
{\em eb10best\/}&$~~c(0,0,0,0)$ & 0 & 0 & 0 & 0 \\
&$~~c(0,0,1,1)$ & 79 & 188 & 1 & 30 \\
&$~~c(1,1,0,0)$ & 68 & 0 & 202 & 31 \\
&$~~c(1,1,1,1)$ & 7793 & 105 & 87 & 223 \\
&$~~other$ &      1160 & 7 & 10 & 16 \\
 \hline
 {\em SAM\/}&$~~c(0,0,0,0)$ & 9095 &  209 &  236 &   193 \\
&$~~c(0,0,1,1)$ &  0  &   7  &   0  &   6 \\
&$~~c(1,1,0,0)$ &  0  &   0  &   0 &    0  \\
&$~~c(1,1,1,1)$ &  0  &   0 &    0  &   0  \\
&$~~other$ &       5  &  84 &   64 &  101 \\
 \hline
\end{tabular}}
\end{center}
\end{table}

\begin{table}[H]
\caption{Confusion matrix for simulation 3.  The column labels indicate the
true underlying patterns and the row labels represent the learned configurations.} \label{pdcCorMotif3}
\begin{center}
\scalebox{0.8}{
\begin{tabular}{c|rrrrrc}
\hline Method & Motif pattern & \multicolumn{1}{c}{$Motif1$}&
\multicolumn{1}{c}{$Motif2$} & \multicolumn{1}{c}{$Motif3$}&
\multicolumn{1}{c}{$Motif4$} & \multicolumn{1}{c}{$Motif5$} \\
\hline
{\em CorMotif\/}& $~~Motif1$ &9189 & 28 & 48 & 50 & 4\\
&$~~Motif2$ &0 & 68 & 0 & 0 & 4 \\
&$~~Motif3$ &0 & 1 & 65 & 0 & 5 \\
&$~~Motif4$ &0 & 2 & 0 & 97 & 6 \\
&$~~Motif5$ &0 & 0 & 0 & 0 & 27 \\
&$~~other$ &11 & 101 & 87 & 53 & 154 \\
\hline
{\em separate limma\/}  &$~~Motif1$ & 9076 & 24 & 36 & 43 & 3 \\
&$~~Motif2$ &  0 & 2 & 0 & 0 & 0 \\
&$~~Motif3$ &0 & 0 & 2 & 0 & 0 \\
&$~~Motif4$ &0 & 0 & 0 & 3 & 1 \\
&$~~Motif5$ &0 & 0 & 0 & 0 & 0 \\
&$~~other$ &124 & 174 & 162 & 154 & 196 \\
\hline
{\em  all concord\/} & $~~Motif1$ &9200 & 96 & 117 & 94 & 5 \\
&$~~Motif2$ &0 & 0 & 0 & 0 & 0 \\
&$~~Motif3$ &0 & 0 & 0 & 0 & 0 \\
&$~~Motif4$ &0 & 0 & 0 & 0 & 0 \\
&$~~Motif5$ & 0 & 104 & 83 & 106 & 195 \\
&$~~other$ &0 & 0 & 0 & 0 & 0 \\
\hline
{\em full motif\/} &$~~Motif1$ & 9185 & 28 & 46 & 49 & 4 \\
&$~~Motif2$ &0 & 63 & 0 & 0 & 3 \\
&$~~Motif3$ &0 & 0 & 51 & 0 & 4 \\
&$~~Motif4$ &0 & 2 & 0 & 89 & 3 \\
&$~~Motif5$ & 0 & 0 & 0 & 0 & 14 \\
&$~~other$ &15 & 107 & 103 & 62 & 172 \\
\hline
{\em eb1\/} &$~~Motif1$ & 748 & 0 & 1 & 1 & 0 \\
&$~~Motif2$ & 273 & 2 & 0 & 0 & 0 \\
&$~~Motif3$ &4 & 0 & 1 & 0 & 0 \\
&$~~Motif4$ &47 & 0 & 0 & 0 & 0 \\
&$~~Motif5$ &1239 & 157 & 149 & 170 & 183 \\
&$~~other$ &6889 & 41 & 49 & 29 & 17 \\
\hline
{\em SAM\/} &$~~Motif1$ & 9200&  139&  170 & 165 & 134 \\
&$~~Motif2$ &  0  &  0  &  0  &  0   & 0 \\
&$~~Motif3$ &0  &  0  &  0  &  0   & 0\\
&$~~Motif4$ &0  &  0  &  0  &  0   & 0 \\
&$~~Motif5$ &0  &  0  &  0  &  0   & 0 \\
&$~~other$ &0  & 61 &  30 &  35 &  66 \\
\hline
\end{tabular}}
\end{center}
\end{table}

\begin{table}[H]
\caption{Confusion matrix for simulation 4. The column labels indicate the
true underlying patterns and the row labels represent the learned configurations.} \label{pdcCorMotif4}
\begin{center}
\scalebox{0.8}{
\begin{tabular}{c|rrrrrc}
\hline Method & Motif pattern & \multicolumn{1}{c}{$Motif1$}&
\multicolumn{1}{c}{$Motif2$} & \multicolumn{1}{c}{$Motif3$}&
\multicolumn{1}{c}{$Motif4$} & \multicolumn{1}{c}{$Motif5$} \\
\hline
{\em CorMotif\/}& $~~Motif1$ &9198 & 4 & 5 & 2 & 0 \\
&$~~Motif2$ &0 & 29 & 0 & 0 & 0 \\
&$~~Motif3$ &0 & 0 & 20 & 0 & 0 \\
&$~~Motif4$ &0 & 0 & 0 & 22 & 0 \\
&$~~Motif5$ &0 & 0 & 0 & 0 & 4 \\
&$~~other$ & 2 & 167 & 175 & 176 & 196 \\
\hline
{\em separate limma\/}  &$~~Motif1$ & 8907 & 1 & 3 & 1 & 0 \\
&$~~Motif2$ &  0 & 0 & 0 & 0 & 0 \\
&$~~Motif3$ & 0 & 0 & 0 & 0 & 0 \\
&$~~Motif4$ & 0 & 0 & 0 & 0 & 0 \\
&$~~Motif5$ & 0 & 0 & 0 & 0 & 0 \\
&$~~other$ &293 & 199 & 197 & 199 & 200 \\
\hline
{\em  all concord\/} & $~~Motif1$ &9200 & 58 & 69 & 69 & 0 \\
&$~~Motif2$ & 0 & 0 & 0 & 0 & 0 \\
&$~~Motif3$ & 0 & 0 & 0 & 0 & 0 \\
&$~~Motif4$ & 0 & 0 & 0 & 0 & 0 \\
&$~~Motif5$ & 0 & 142 & 131 & 131 & 200 \\
&$~~other$ & 0 & 0 & 0 & 0 & 0 \\
\hline
{\em  SAM\/} & $~~Motif1$ &9197 &  64 &  66 &  92 &  23  \\
&$~~Motif2$ & 0 & 0 & 0 & 0 & 0 \\
&$~~Motif3$ & 0 & 0 & 0 & 0 & 0 \\
&$~~Motif4$ & 0 & 0 & 0 & 0 & 0 \\
&$~~Motif5$ & 0 & 0 & 0 & 0 & 0 \\
&$~~other$ & 3  &136&  134 & 108 & 177 \\
\hline
\end{tabular}}
\end{center}
\end{table}

\begin{table}[H]
\caption{GEO data used for real data based simulations.}
\label{pdceb10}
\begin{center}
\scalebox{0.7}{\begin{tabular}{cccccc}
\hline Simulation ID & Study ID & GEO Sample Id & GEO series number & Sample No. & Sample type\\
\hline
Simulations 5-10 &1 &GSM366065.CEL - GSM366070.CEL& GSE14668 & 6 & Liver tissue of liver donor\\
Simulations 5-10 &2 &GSM550623.CEL - GSM550628.CEL& GSE22138 & 6 & Uveal Melanoma primary tumor tissue\\
Simulations 5-10 &3 &GSM553482.CEL - GSM553487.CEL & GSE22224 & 6 & Peripheral blood mononuclear cells of healthy volunteer\\
Simulations 5-10 &4 &GSM494634.CEL - GSM494639.CEL & GSE33356 & 6 & Normal lung tissue\\
Simulations 6-7&5 &GSM909644.CEL - GSM909649.CEL & GSE37069 & 6 & Blood samples from controls\\
Simulations 6-7&6 &GSM909650.CEL - GSM909655.CEL & GSE37069 & 6 & Blood samples from controls\\
Simulations 6-7&7 &GSM909656.CEL - GSM909661.CEL & GSE37069 & 6 & Blood samples from controls\\
Simulations 6-7&8 &GSM909662.CEL - GSM909667.CEL & GSE37069 & 6 & Blood samples from controls\\
Simulations 6-7&9 &GSM90968.CEL - GSM909673.CEL & GSE37069 & 6 & Blood samples from controls\\
Simulations 6-7&10 &GSM909674.CEL - GSM909679.CEL & GSE37069 & 6 & Blood samples from controls\\
Simulation 7&11 &GSM376428.CEL - GSM376433.CEL & GSE15061& 6 & Non-leukemia bone marrow samples\\
Simulation 7&12 &GSM376434.CEL - GSM376439.CEL & GSE15061& 6 & Non-leukemia bone marrow samples\\
Simulation 7&13 &GSM376440.CEL - GSM376445.CEL & GSE15061& 6 & Non-leukemia bone marrow samples\\
Simulation 7&14 &GSM376446.CEL - GSM376451.CEL & GSE15061& 6 & Non-leukemia bone marrow samples\\
Simulation 7&15 &GSM376452.CEL - GSM376457.CEL & GSE15061& 6 & Non-leukemia bone marrow samples\\
Simulation 7&16 &GSM376458.CEL - GSM376463.CEL & GSE15061& 6 & Non-leukemia bone marrow samples\\
Simulation 7&17 &GSM376464.CEL - GSM376469.CEL & GSE15061& 6 & Non-leukemia bone marrow samples\\
Simulation 7&18 &GSM376470.CEL - GSM376475.CEL & GSE15061& 6 & Non-leukemia bone marrow samples\\
Simulation 7&19 &GSM376476.CEL - GSM376481.CEL & GSE15061& 6 & Non-leukemia bone marrow samples\\
Simulation 7&20 &GSM376482.CEL - GSM376487.CEL & GSE15061& 6 & Non-leukemia bone marrow samples\\
\hline
\end{tabular}}
\end{center}
\end{table}

\begin{table}[htbp]
\caption{Confusion matrix for simulation 5. The column labels indicate the
true underlying patterns and the row labels represent the learned configurations.}
\begin{center}
\scalebox{0.8}{
\begin{tabular}{c|rrrrc}
\hline Method & Motif pattern & \multicolumn{1}{c}{$c(0,0,0,0)$} &
\multicolumn{1}{c}{$c(0,0,1,1)$} &
\multicolumn{1}{c}{$c(1,1,0,0)$} &\multicolumn{1}{c}{$c(1,1,1,1)$}\\
\hline
{\em CorMotif\/} &$~~c(0,0,0,0)$ &53670 & 108 & 164 & 20\\
&$~~c(0,0,1,1)$ &  6 & 286 & 0 & 18 \\
&$~~c(1,1,0,0)$ &29 & 0 & 200 & 6 \\
&$~~c(1,1,1,1)$&  0 & 0 & 0 & 31 \\
&$~~other$ & 70 & 6 & 36 & 25 \\
\hline
{\em separate limma\/} & $~~c(0,0,0,0)$ & 53615 & 121 & 171 & 24 \\
&$~~c(0,0,1,1)$ &  0 & 79 & 0 & 8 \\
&$~~c(1,1,0,0)$ &  0 & 0 & 46 & 3 \\
&$~~c(1,1,1,1)$ &  0 & 0 & 0 & 1 \\
&$~~other$ & 160 & 200 & 183 & 64 \\
\hline
{\em all concord\/} & $~~c(0,0,0,0)$  & 53748 & 187 & 255 & 26 \\
& $~~c(0,0,1,1)$ & 0 & 0 & 0 & 0 \\
& $~~c(1,1,0,0)$ & 0 & 0 & 0 & 0 \\
&$~~c(1,1,1,1)$ &  27 & 213 & 145 & 74 \\
&$~~other$ & 0 & 0 & 0 & 0 \\
 \hline
{\em full motif \/}&$~~c(0,0,0,0)$ & 53671 & 108 & 165 & 20 \\
&$~~c(0,0,1,1)$ & 5 & 286 & 0 & 18 \\
&$~~c(1,1,0,0)$ & 30 & 0 & 201 & 6 \\
&$~~c(1,1,1,1)$ & 0 & 0 & 1 & 36 \\
&$~~other$ &  69 & 6 & 33 & 20 \\
\hline
{\em eb1\/} &$~~c(0,0,0,0)$ & 49817 & 190 & 188 & 23 \\
&$~~c(0,0,1,1)$ & 161 & 103 & 0 & 12 \\
&$~~c(1,1,0,0)$ &  244 & 0 & 66 & 8 \\
&$~~c(1,1,1,1)$ & 11 & 0 & 0 & 7 \\
&$~~other$ & 3542 & 107 & 146 & 50 \\
 \hline
{\em eb10best\/}&$~~c(0,0,0,0)$ & 51731 & 109 & 125 & 36 \\
&$~~c(0,0,1,1)$ & 5 & 232 & 0 & 6 \\
&$~~c(1,1,0,0)$ & 12 & 0 & 169 & 4 \\
&$~~c(1,1,1,1)$ & 0 & 0 & 0 & 16 \\
&$~~other$ & 2027 & 59 & 106 & 38 \\
 \hline
 {\em SAM\/}&$~~c(0,0,0,0)$ & 53773&   283 &  398  &  83  \\
&$~~c(0,0,1,1)$ & 0 &   0  &  0   & 0 \\
&$~~c(1,1,0,0)$ & 0 &   0  &  0   & 0 \\
&$~~c(1,1,1,1)$ & 0 &   0  &  0   & 0\\
&$~~other$ &  2 & 117  &  2 &  17\\
\hline
\end{tabular}}
\end{center}
\end{table}

\begin{table}[H]
\caption{Confusion matrix for simulation 6.  The column labels indicate the
true underlying patterns and the row labels represent the learned configurations.} \label{pdcCorMotif3}
\begin{center}
\scalebox{0.8}{
\begin{tabular}{c|rrrrrc}
\hline Method & Motif pattern & \multicolumn{1}{c}{$Motif1$}&
\multicolumn{1}{c}{$Motif2$} & \multicolumn{1}{c}{$Motif3$}&
\multicolumn{1}{c}{$Motif4$} & \multicolumn{1}{c}{$Motif5$} \\
\hline
{\em CorMotif\/}& $~~Motif1$& 53600 & 15 & 11 & 15 & 1 \\
&$~~Motif2$& 0 & 169 & 0 & 1 & 4\\
&$~~Motif3$& 4 & 1 & 147 & 0 & 2\\
&$~~Motif4$& 1 & 3 & 0 & 178 & 7\\
&$~~Motif5$& 0 & 1 & 0 & 1 & 170\\
&$~~other$& 270 & 11 & 42 & 5 & 16\\
 \hline
{\em separate limma\/}&$~~Motif1$& 53340 & 21 & 12 & 22 & 5\\
&$~~Motif2$ & 0 & 16 & 0 & 0 & 4\\
&$~~Motif3$& 0 & 0 & 14 & 0 & 2\\
&$~~Motif4$& 0 & 0 & 0 & 17 & 1\\
&$~~Motif5$& 0 & 0 & 0 & 0 & 0\\
&$~~other$ & 535 & 163 & 174 & 161 & 188\\
   \hline
{\em all concord\/}&$~~Motif1$& 43 & 36 & 49 & 4 \\
&$~~Motif2$& 0 & 0 & 0 & 0 & 0 \\
&$~~Motif3$& 0 & 0 & 0 & 0 & 0\\
&$~~Motif4$& 0 & 0 & 0 & 0 & 0\\
&$~~Motif5$& 17 & 157 & 164 & 151 & 196\\
&$~~other$& 0 & 0 & 0 & 0 & 0\\
   \hline
{\em full motif \/}&$~~Motif1$& 53578 & 15 & 11 & 13 & 1 \\
&$~~Motif2$& 0 & 156 & 0 & 0 & 2 \\
&$~~Motif3$& 3 & 0 & 146 & 0 & 1 \\
&$~~Motif4$& 1 & 2 & 0 & 166 & 4\\
&$~~Motif5$& 0 & 0 & 0 & 0 & 136 \\
&$~~other$& 293 & 27 & 43 & 21 & 56\\
   \hline
{\em eb1\/} & $~~Motif1$& 47986 & 24 & 14 & 18 & 0\\
& $~~Motif2$ & 3 & 47 & 0 & 0 & 5\\
& $~~Motif3$& 23 & 1 & 42 & 0 & 1\\
& $~~Motif4$ & 10 & 0 & 0 & 69 & 1\\
& $~~Motif5$& 3 & 0 & 0 & 0 & 38\\
& $~~other$& 5850 & 128 & 144 & 113 & 155\\
\hline
{\em SAM\/} & $~~Motif1$& 53851 & 120 & 138&  116 &  89\\
& $~~Motif2$ &  0  &  0 &   0 &   0  &  0\\
& $~~Motif3$& 0  &  0 &   0 &   0  &  0\\
& $~~Motif4$ & 0  &  0 &   0 &   0  &  0\\
& $~~Motif5$& 0  &  0 &   0 &   0  &  0\\
& $~~other$&  24 &  80&   62&   84 & 111\\
\hline
\end{tabular}}
\end{center}
\end{table}

\begin{table}[H]
\caption{Confusion matrix for simulation 7.  The column labels indicate the
true underlying patterns and the row labels represent the learned configurations.} \label{pdcCorMotif3}
\begin{center}
\scalebox{0.8}{
\begin{tabular}{c|rrrrrc}
\hline Method & Motif pattern & \multicolumn{1}{c}{$Motif1$}&
\multicolumn{1}{c}{$Motif2$} & \multicolumn{1}{c}{$Motif3$}&
\multicolumn{1}{c}{$Motif4$} & \multicolumn{1}{c}{$Motif5$} \\
\hline
{\em CorMotif\/}& $~~Motif1$&52442 & 3 & 5 & 4 & 1\\
& $~~Motif2$& 6 & 188 & 0 & 0 & 1 \\
& $~~Motif3$& 10 & 0 & 156 & 0 & 0 \\
& $~~Motif4$& 5 & 0 & 0 & 187 & 10 \\
& $~~Motif5$& 0 & 0 & 0 & 0 & 165\\
& $~~other$& 1412 & 9 & 39 & 9 & 23\\
  \hline
{\em separate limma \/}& $~~Motif1$&51999 & 7 & 24 & 5 & 4 \\
& $~~Motif2$ & 0 & 0 & 0 & 0 & 0\\
& $~~Motif3$& 0 & 0 & 0 & 0 & 0 \\
& $~~Motif4$ & 0 & 0 & 0 & 0 & 0 \\
& $~~Motif5$& 0 & 0 & 0 & 0 & 0 \\
& $~~other$ & 1876 & 193 & 176 & 195 & 196\\
   \hline
{\em all concord \/}& $~~Motif1$&53859 & 27 & 49 & 18 & 3 \\
& $~~Motif2$& 0 & 0 & 0 & 0 & 0\\
& $~~Motif3$& 0 & 0 & 0 & 0 & 0\\
& $~~Motif4$& 0 & 0 & 0 & 0 & 0\\
& $~~Motif5$& 16 & 173 & 151 & 182 & 197\\
& $~~other$ & 0 & 0 & 0 & 0 & 0\\
   \hline
{\em SAM \/}& $~~Motif1$&53812 & 108 & 145 & 110 & 100 \\
& $~~Motif2$& 0 & 0 & 0 & 0 & 0\\
& $~~Motif3$& 0 & 0 & 0 & 0 & 0\\
& $~~Motif4$& 0 & 0 & 0 & 0 & 0\\
& $~~Motif5$& 0 & 0 & 0 & 0 & 0\\
& $~~other$ & 63 &  92 &  55 &  90 & 100\\
   \hline
\end{tabular}}
\end{center}
\end{table}

\begin{table}
\caption{Ranks of known SHH target genes by each method in the SHH analysis.}
\begin{center}
\scalebox{1}{
\begin{tabular}{crrrrrrrc}
\hline Gene name & Analysis Method& Study 1& Study 2& Study 3& Study 4 & Study 5& Study 6& Study 7\\
\hline
Gli1 & {\em separate limma  \/}& 6 &7& 16&9&7&1369&515 \\
&{\em CorMotif \/}& 5&   6&   7&   7&   6& 930& 324 \\
& {\em all concord \/} & 9 &9& 9&9&9&9&9 \\
&{\em full motif \/} &5  & 7  & 7 &  4&   5 &809& 308\\
&{\em SAM \/}&  7  &  6  & 17 &   9 &  10& 1627 & 583\\
&{\em eb1 \/}& 33396 &   25 &   36 &   24  &  24&  1828  & 720\\
\hline
Ptch1&{\em separate limma \/}&7 &  19 &    4  &   4   &  2 &  783  &  19\\
&{\em CorMotif \/}&6 &  20  &  8 &   4  &  3 &  495 &   12\\
&{\em all concord  \/}&5  &  5  &  5 &   5 &   5  &  5  &  5\\
&{\em full motif \/}&  7  & 16 &   4 &   3  &  2  & 409  &  14\\
&{\em SAM \/}&  6  & 18   &  5   &  4   &  2 &  964  &  25\\
&{\em eb1 \/}&13455  &  8  &  6 &   9   & 4 & 1464 &  289\\
\hline
Ptch2&{\em separate limma \/}&273&  607& 9996& 1527 & 458& 2530 & 117\\
&{\em CorMotif \/}&140&  437&  462 & 356  &264& 1848 &  69\\
&{\em all concord  \/}&40& 40& 40& 40& 40& 40& 40\\
&{\em full motif \/}&145 & 450&  482&  285&  256 &1686  & 70\\
&{\em SAM \/}& 303 & 630& 9066 &1431 & 468 &2488  & 95\\
&{\em eb1 \/}&7331 & 579 & 838 & 727&  433 & 418 & 161\\
\hline
Hhip&{\em separate limma \/}&105  & 25 &  31 &  580&  2964& 13452  &  6\\
&{\em CorMotif \/}&61&   19 &  27&  264 & 652&  9259 &   2\\
&{\em all concord  \/}&22 &  22 &  22  & 22  & 22  & 22  & 22\\
&{\em full motif \/}&58  & 22&   28 & 249 & 632 & 8529  &  2\\
&{\em SAM \/}&  107  & 24 &  20  & 597 & 2903& 16223   & 7\\
&{\em eb1 \/}&6111 &  32  & 10 & 353&  326 & 7462 & 131\\
\hline
Rab34&{\em separate limma \/}& 927&   553 &  299&   577&   396&  15782&   241\\
&{\em CorMotif \/}&324 & 401&  164&  176 & 261 &10418 & 150\\
&{\em all concord \/} & 160 & 160 & 160 & 160&  160 & 160 & 160\\
&{\em full motif \/}& 386 & 372&  139 & 194 & 274 & 9546&  151\\
&{\em SAM \/}& 953 & 613 & 450 & 619 & 430 &15923  &171\\
&{\em eb1 \/}&1371 &1333& 1042& 1130 &1074 &12564& 1019\\
\hline
Hand2&{\em separate limma \/}& 34351 &11862 & 6647 & 6061&   196 &20672& 44939\\
&{\em CorMotif \/}&3601  &3394 & 2794 & 1036 &  544 &13371 &17909\\
&{\em all concord  \/}&4987& 4987& 4987 &4987 &4987& 4987& 4987\\
&{\em full motif \/}&3327 & 3021 & 2460  & 917  & 550 &12585 &14457\\
&{\em SAM \/}& 34455& 12375 & 8381 & 6582 &  207 &22592& 44945\\
&{\em eb1 \/}&28270 & 2191&  3040 & 1650 &  571& 23269& 33457\\
\hline
Hoxd13&{\em separate limma \/}&  6805 & 7572 & 1893& 10644 &  12 &26047 & 9676\\
&{\em CorMotif \/}&1990 &2371 &1746 &1223  & 93& 15204 & 5734\\
&{\em all concord  \/}&933 & 933 & 933&  933&  933 & 933&  933\\
&{\em full motif \/}&1943 &2490 &1246 &1064  & 88 &14041 & 4722\\
&{\em SAM \/}&  6724 & 7763&  2684 &10553 &  12 &27578 & 8579\\
&{\em eb1 \/}& 6919 & 804 & 696 & 641  & 14 &26742 &12464\\
\hline
\end{tabular}}
\end{center}
\label{shhtopgene1}
\end{table}